\documentclass[screen, sigplan, nonacm]{acmart}     
\usepackage[utf8]{inputenc}
\usepackage{microtype}
\usepackage{nicefrac}
\usepackage{qtree}

\usepackage{amsmath, amsthm, amssymb, amsfonts, mathtools}
\usepackage{bm} 
\usepackage{bbm} 
\usepackage{csquotes}

\usepackage{tikz}
\usetikzlibrary{quantikz2}
\usepackage{physics}

\usepackage{xcolor}
\definecolor{custom1}{HTML}{023047} 
\definecolor{custom2}{HTML}{ffb703} 
\definecolor{custom3}{HTML}{fb8500} 
\definecolor{orangepeel}{rgb}{1.0, 0.62, 0.0}

\usepackage{graphicx}
\graphicspath{{figures/}}
\usepackage{float}
\usepackage{booktabs, makecell, tabularx}
\usepackage{multirow}
\usepackage{subcaption}
\usepackage{adjustbox}
\usepackage{array} 

\usepackage[linesnumbered,ruled,vlined]{algorithm2e}
\usepackage{listings}

\PassOptionsToPackage{hyphens}{url}
\usepackage{hyperref}
\AtBeginDocument{\hypersetup{
    colorlinks=true,
    linkcolor=violet,
    urlcolor=blue,
    citecolor=orangepeel,
}}
\usepackage[capitalise]{cleveref}

\usepackage{pgfplots}
\pgfplotsset{compat=1.16}

\usepackage{xspace}
\usepackage{enumerate}
\usepackage{comment}
\usepackage{pifont}
\usepackage{stmaryrd} 

\newcommand{\Qubit}{\mathcal{Q}}
\newcommand{\Ancilla}{\mathcal{A}}
\newcommand{\qcode}[3]{\llbracket #1,#2,#3 \rrbracket}

\newcommand{\curly}[1]{\left\{ #1 \right\}}

\newcommand{\mbf}[1]{\mathbf{#1}}
\newcommand{\mcl}[1]{\mathcal{#1}}

\usepackage{todonotes}

\allowdisplaybreaks

\DeclareUnicodeCharacter{2009}{\,}

\begin{document}

\title{Quantum Error Detection For Early Term Fault-Tolerant Quantum Algorithms}
\author{Tom Ginsberg}
\email{tom@beit.tech}
\affiliation{
    \institution{BEIT Inc.}
    \country{}
}
\author{Vyom Patel}
\email{vyom.patel@uwaterloo.ca}
\affiliation{
    \institution{University of Waterloo \& BEIT Inc.}
    \country{}
}

\begin{abstract}
Quantum error detection (QED) offers a promising pathway to fault tolerance in near-term quantum devices by balancing
error suppression with minimal resource overhead. However, its practical utility hinges on optimizing design 
parameters—such as syndrome measurement frequency—to avoid diminishing returns from detection overhead. In this work,
we present a comprehensive framework for fault-tolerant compilation and simulation of quantum algorithms using the
$\qcode{n}{n-2}{2}$ codes, which enables low-qubit-overhead error detection and a simple nearly fault-tolerant universal
set of operations. We demonstrate and analyze our pipeline with a purely statistical interpretation and through the
implementation of Grover's search algorithm. Our results are used to answer the question
\textit{is quantum error detection a worthwhile avenue for early-term fault tolerance, and if so how can we get the most out of it?}
Simulations under the circuit-level noise model reveal that finding optimal syndrome schedules improve algorithm
success probabilities by an average of $6.7\times$ but eventual statistical limits from post-selection in
noisy/resource-limited regimes constrain scalability. Furthermore, we propose a simple data-driven approach to predict fault tolerant compilation parameters, such as optimal syndrome schedules, and expected fault tolerant performance gains based on circuit and noise features. These results provide actionable guidelines for implementing QED
in early-term quantum experiments and underscore its role as a pragmatic, constant-overhead error mitigation layer for
shallow algorithms. To aid in further research, we release all simulation data computed for this work and provide an experimental QED compiler at \href{https://codeqraft.xyz/qed}{codeqraft.xyz/qed}. 
\end{abstract}

\maketitle

\section{Introduction}\label{sec:introduction}

\begin{figure*}[!htb]
    \centering
    \begin{minipage}[t]{0.49\linewidth}
        \centering
        \includegraphics[width=\linewidth]{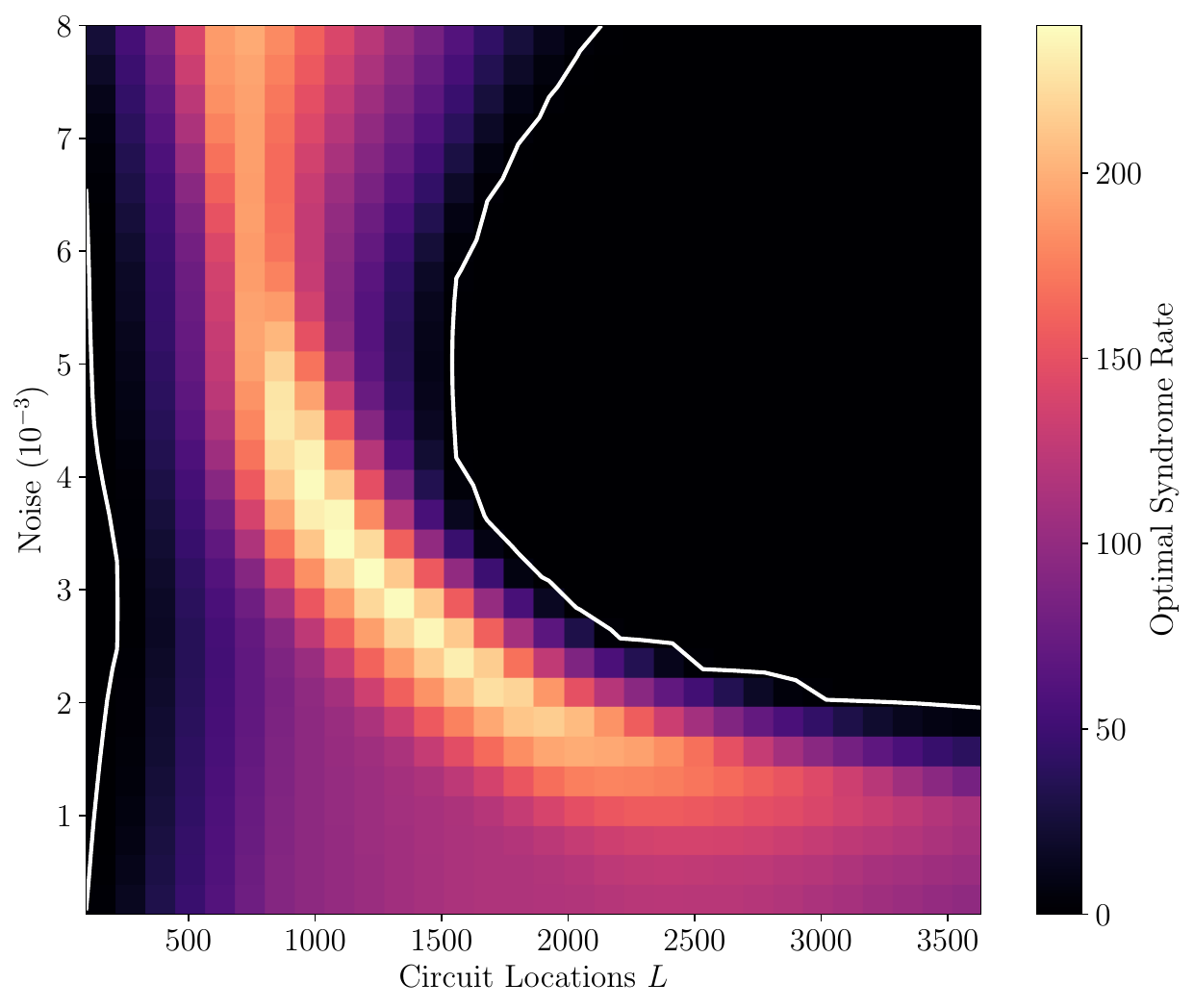}
    \end{minipage}
    \hfill
    \begin{minipage}[t]{0.49\linewidth}
        \centering
        \includegraphics[width=\linewidth]{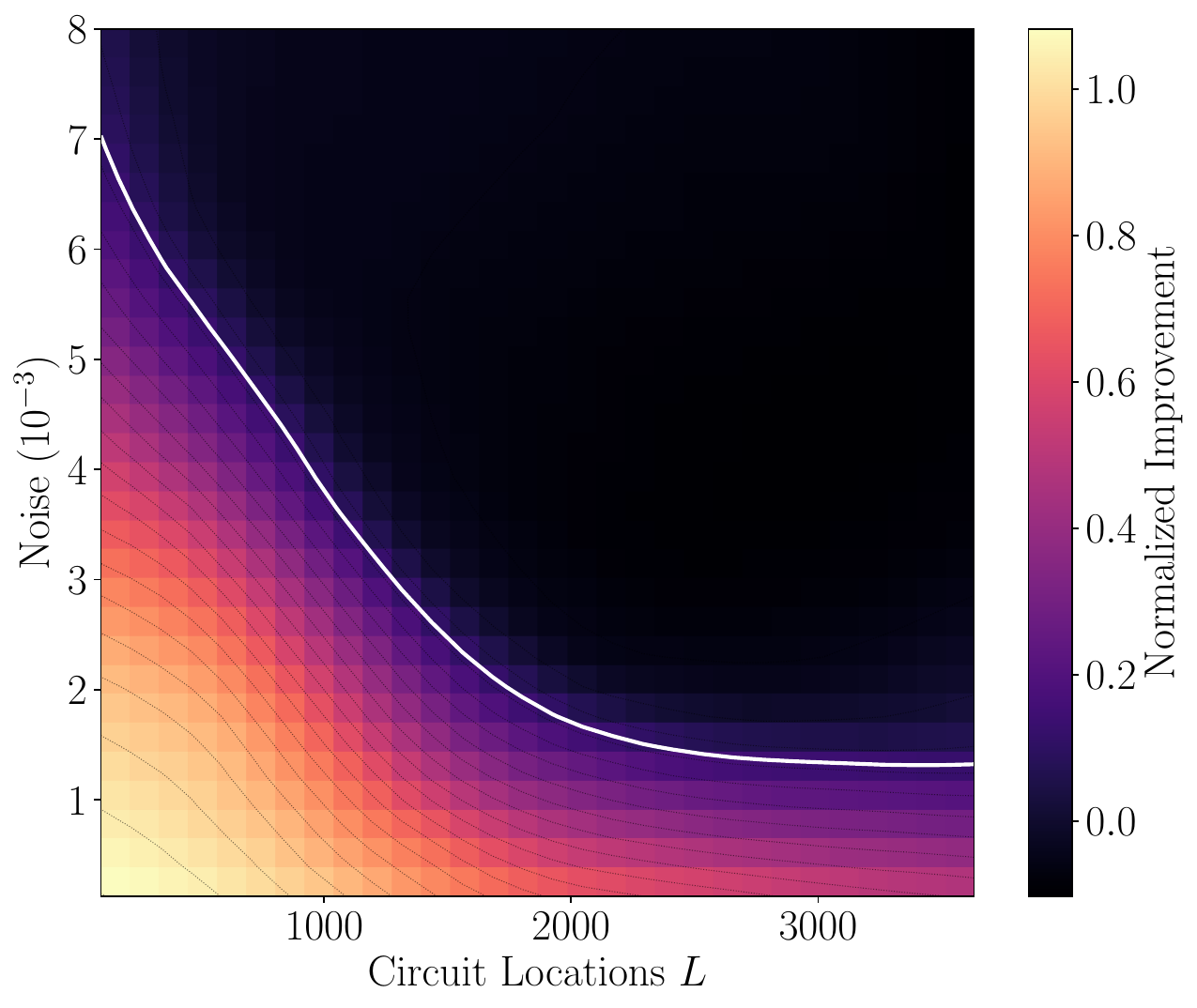}
    \end{minipage}
	\caption{Heatmaps showing experimental results for Grover's algorithm with a 
	single marked state as functions of circuit locations ($L$) and noise model parameter ($p$). 
	 (\textit{Left: Optimal Syndrome Rate}) The rate of syndrome measurements per logical operation 
	 that maximizes the probability of measuring the marked state. A white boundary indicates configurations 
	 where the optimal syndrome count is one, signifying no utility from adding mid-circuit syndrome measurements 
	 beyond the final readout. (\textit{Right: Normalized Success rate improvement}), 
	 The ratio of success probabilities between the encoded implementation $p_{\text{enc}}$ using quantum error 
	 detection (with the optimal syndrome count) and a bare implementation $p_{\text{bare}}$ without error detection under 
	 identical noise conditions normalized up to $p_{\text{ideal}}$ the noiseless success probability: $\nicefrac{p_{\text{enc}} - p_{\text{bare}}}{p_{\text{ideal}} -p_{\text{bare}}}$. The white boundary corresponds to a ratio where adding fault tolerance does not improve performance.}    \label{fig:combined-heatmaps}
\end{figure*}

Quantum error correction (QEC) has emerged as a cornerstone in the pursuit of scalable quantum computing, aiming to
protect quantum information from decoherence and operational errors. The foundational theory of quantum error correction,
developed over the past few decades provides a range of techniques to mitigate errors through redundantly encoding
quantum information across multiple physical qubits. While the ultimate goal is to achieve fully fault-tolerant quantum
computation with logical error rates below any conceivable noise threshold, early fault-tolerant quantum computing
(EFTQC), focuses on demonstrating meaningful quantum algorithms with manageable resource overheads. This paradigm shift
has been driven by recent advances in quantum computing hardware where several dozen to hundreds of densely connected and
high-fidelity qubits are available today or expected to be in the near future~\cite{decross2024computational, bluvstein2024logical}.

\textit{Quantum error detection} (QED) has gained renewed attention
~\cite{iceberg, PRXQuantum.5.020101, chen2024tailoring, hangleiter2024faulttolerant, hong2024entangling,
wang2024fault, yamamoto2024demonstrating, he2024performance} in this context due to its potential for lower qubit
overheads and simpler circuit requirements. Unlike full error correction, which involves both detecting and correcting
errors, using mid-circuit classical feedback, quantum error detection focuses solely on identifying and selecting out
errors as a post-processing step making it also fall into the category of error mitigation. This approach can be 
advantageous for early demonstrations of fault tolerance, particularly in architectures where qubit resources and 
real-time feedback are limited. The progress in the early fault-tolerant demonstrations has been facilitated by several key 
technological developments. For instance, neutral atom~\cite{bluvstein2024logical, norcia2024iterative} and trapped ion 
processors~\cite{chen2023benchmarking, moses2023race, decross2024computational}, which offer high-fidelity qubit 
operations and lower connectivity constraints, present new avenues for resource-efficient quantum error correction 
beyond the quadratic overhead of the traditional surface code~\cite{dennis2002topological, acharya2024quantum}. These 
platforms allow for flexible qubit arrangements and tunable interactions, reducing the complexity of implementing 
error-detecting codes and other fault-tolerant protocols. This flexibility is crucial for designing circuits that 
maximize computational throughput while minimizing the quantum resources required for error correction.

Moreover, the compilation of quantum algorithms into fault-tolerant circuits has become an active area of research.
The process involves translating logical quantum circuits into sequences of fault-tolerant gadgets based on the algebraic
structure specific the error-correcting code. When limited to planar connectivity, the surface code~\cite{dennis2002topological}
enables universal fault tolerance at a quadratic space complexity cost in the Clifford+T gate set via lattice 
surgery~\cite{horsman2012surface} and magic state distillation~\cite{gidney2024magic}. However, denser connectivity
leads to considerably more open-ended design space for lower time/space overhead and even distillation-free 
approaches~\cite{Paetznick_2013, anderson2014fault, jochym2014using, goto2024manyhypercube}. In general logical
compilation is a huge and rapidly growing space highlighting the importance of both hardware-specific and 
algorithm-specific considerations in achieving practical fault tolerance, an idea often referred to as quantum error 
correction \emph{co-design} in the community. Quantum error detection, in particular, due to its low overheads, 
represents a potentially promising avenue for demonstrating meaningful fault-tolerant quantum algorithms in the near
term. Beyond its benefit for demonstrations, QED offers a simple paradigm to explore fault-tolerant design trade-offs
between computational complexity and noise resilience, providing valuable lessons in a road map for scaling up to more
complex universal error correction.

In this work, we present a fault-tolerant compilation pipeline and simulation study for Grover's search 
algorithm~\cite{grover1996fast} utilizing $\qcode{n}{n-2}{2}$ quantum error-detecting 
codes~\cite{steane1996simple, gottesman1998theory}. By examining the impact of various circuit and QEC design 
parameters—such as the dimension of the search space, the number of marker-diffusion iterations, and the frequency of 
syndrome measurements, we aim to delineate the regime where quantum error detection provides computational advantages.

This paper contributes to the ongoing discourse on scalable and efficient quantum algorithms by offering both theoretical
and experimental insights into the utility of quantum error detection. We argue both with theory and supporting 
experiments, that with careful tuning of design parameters, progress can be made with quantum error detection, but 
statistical limits clearly prevent its use case as a truly scalable fault-tolerant paradigm.

\section{Quantum Error Correction}\label{sec:quantum-error-correction}
\subsection{Stabilizer Codes}\label{subsec:stabilizer-codes}
An $\qcode{n}{k}{d}$ quantum error-correcting code uses $n$ physical qubits to encode $k < n$ logical qubits with a 
code distance of $d$. A general code with distance $d$ allows the system to correct up to $\lfloor(d-1)/2\rfloor$ and 
detect up to $d-1$ individual qubit errors. The defining feature of a stabilizer code is its stabilizer group, a group 
generated by $r = n - k$ commuting $n$-qubit Pauli operators $G$ such that $-I\notin \mathrm{span}(G)$. The logical 
space of the code, or codespace, is determined by the simultaneous $+1$ eigenstates of these stabilizers.

\subsection{Quantum Error Detection: $\qcode{n}{n-2}{2}$ Codes}\label{subsec:quantum-error-detection}
A promising approach for early-term fault-tolerant quantum algorithms is the use of quantum error detection. This method
offers low encoding and compilation overhead while enabling the detection and subsequent removal of sample results known 
to have errors. The $\qcode{n}{n-2}{ 2}$ code, more recently named the \textit{iceberg code}, is a well-known 
error-detecting code noted for its high encoding rate and support for fault-tolerant Clifford group operations
~\cite{Linke_2017, flags, iceberg, goto2024manyhypercube}. Very recent works have shown the iceberg code can be used to
effectively implement Trotter circuits \cite{chen2024tailoring} and QAOA \cite{he2024performance} for small circuit sizes.

The $\qcode{n}{n-2}{2}$ code, defined for even $n \geq 4$, is generated by the stabilizers $X^{\otimes n}$ and 
$Z^{\otimes n}$ meaning logical states are defined by a global parity check in the $X$ and $Z$ basis. The common choice 
is to set the logical zero state to the $n$-qubit GHZ state, $|\overline{0}\rangle^{\otimes k} \coloneqq \nicefrac{1}{\sqrt{2}} 
\left(\ket{0}^{\otimes n} + \ket{1}^{\otimes n}\right)$ (with $k\coloneqq n-2$) which can be prepared fault-tolerantly with a repeat until
success loop using $n+1$ CNOT total gates and one measurement of a simple weight two parity check (see \cref{alg:ft-zero-state}). 

Logical Pauli operators are set to $\overline{X}_i = X_1 X_{i+1}$ and $\overline{Z}_i = Z_{i+1}Z_{n}$ for 
$i \in \{1,\dots, n\}$. Via simplification over the stabilizer group, all qubit operators $\overline{X}^{\otimes k}$ 
and $\overline{Z}^{\otimes k}$ can be implemented as weight two operators $X_1 X_{n}$ and $Z_1 Z_{n}$ respectively. In 
fact, it becomes easy to see that any logical $\overline{X}$ or $\overline{Z}$ operator on a subset of logical qubits 
will always be realizable by a weight two physical operator of the same type (see \cref{alg:logical-paulis}). The 
$\qcode{n}{n-2}{2}$ is also notable for its swap-transversal Hadamard on all $\overline{H}^{\otimes k}$ gate that is 
realized by a physical $H^{\otimes n}$ gate followed by a $\text{SWAP}_{1 \leftrightarrow n}$
~\cite{wang2023faulttolerant, goto2024manyhypercube}. Note that on dense hardware topologies swap operations become 
trivial as qubits can be renumbered arbitrarily. See \cref{sec:subroutines} for full subroutines of fault-tolerant state 
preparation, fault-tolerant measurement gadgets, and \cref{fig:gate_compilation} for circuits demonstrating several 
logical gate implementations.

\section{Fault Tolerant Compilation}\label{sec:fault-tolerant-compilation}
Fault-tolerant compilation (FTC) refers to the process of transforming logical quantum algorithms into fault-tolerant 
circuits that can be executed on noisy quantum hardware while maintaining computational integrity. While quantum 
error-correcting codes form the foundation of FTC, a complete fault-tolerant protocol must also include precise 
constructions for fault-tolerant state preparation, logical gate operations, syndrome extraction, and destructive 
measurements, all of which must preserve the code distance. The overhead induced by FTC varies significantly depending 
on the choice of error-correcting code, the connectivity, and native gate set of the hardware, and the computational 
demands of the target algorithm. Effective FTC requires a co-design approach that optimizes these factors collectively, 
an idea emphasized by Daniel Gottesman as early as 2013~\cite{gottesman2013fault}.

\subsection{Universal Compilation for $\qcode{n}{n-2}{2}$ Codes}\label{subsec:compilation}

In~\cite{iceberg}, the authors propose a low-overhead method for realizing a \textit{nearly fault-tolerant} universal 
gate set on $\qcode{n}{n-2}{2}$ codes, leveraging
\[
	\text{RZZ}_{i,j}(\theta) := \exp\left(- i \frac{\pi \theta}{2} Z_i \otimes Z_j\right)
\]
gates, which are native to the Quantinuum H2 ion-trap processor~\cite{Moses_2023}.
\smallbreak\noindent
To summarize their process:
\begin{enumerate}
	\item \emph{Transpile} the target logical unitary to the gate set 
    $\text{RX}_i (\theta) := \exp\left(- i \frac{\pi \theta}{2} X_i\right)$, $\text{RZ}_i (\theta) := \exp\left(- i \frac{\pi \theta}{2} Z_i\right)$,
    and $\text{RZZ}_{i,j} (\theta)$. This step can be performed using transpilers like \texttt{qiskit}
    ~\cite{javadi2024quantum} or \texttt{tket}~\cite{sivarajah2020t}.

	\item \emph{Replace} logical gates with physical counterparts, using the following mappings for Pauli gates:
	      \begin{align}
		       & \overline{\text{RX}}_i \rightarrow \text{RXX}_{1, i+1},       \\
		       & \overline{\text{RZ}}_i \rightarrow \text{RZZ}_{i+1, n},       \\
		       & \overline{\text{RZZ}}_{i,j} \rightarrow \text{RZZ}_{i+1, j+1}
	      \end{align}

	\item \emph{Convert and Squash} the $\text{RXX}_{1, i+1}$ gate:
	      \begin{align}
		       & \text{RXX}_{1, i+1} \rightarrow (H_1 \otimes H_{i+1}) 
		        \text{RZZ}_{1, i+1} (H_1 \otimes H_{i+1})
	      \end{align}
	      Afterward, repeated single-qubit gates (e.g., $H \otimes H$) can be eliminated or combined into arbitrary 
          single-qubit rotations.
\end{enumerate}

\begin{figure*}[!htbp]
    \centering
    \begin{subfigure}[b]{0.16\textwidth}
        \begin{quantikz}[row sep={0.7cm,between origins}]
            \lstick{$1$} & \gate{X} & \\
            \lstick{$i+1$} & \gate{X} & \\
            \lstick{$n$} &	    & \\
        \end{quantikz}
        \caption{$\overline{X}_i$}
    \end{subfigure}
    \hfill
    \begin{subfigure}[b]{0.16\textwidth}
        \begin{quantikz}[row sep={0.7cm,between origins}]
            \lstick{$1$} & \gate{X} & \\
            \lstick{$i+1$} & \gate{Y} & \\
            \lstick{$n$} & \gate{Z} & \\
        \end{quantikz}
        \caption{$\overline{Y}_i$}
    \end{subfigure}
    \hfill
    \begin{subfigure}[b]{0.16\textwidth}
        \begin{quantikz}[row sep={0.7cm,between origins}]
            \lstick{$1$} &	    & \\
            \lstick{$i+1$} & \gate{Z} & \\
            \lstick{$n$} & \gate{Z}	 & \\
        \end{quantikz}
        \caption{$\overline{Z}_i$}
    \end{subfigure}
    \hfill
    \begin{subfigure}[b]{0.27\textwidth}
        \begin{quantikz}[row sep={0.7cm,between origins}]
            \lstick{$1$} &		&	   &	     & \\
            \lstick{$i+1$} &	    & \ctrl{1} & \gate{S}& \\
            \lstick{$n$} & \gate{S} & \phase{} &		 & \\
        \end{quantikz}
        \caption{$\overline{S}_i$}
    \end{subfigure}
    \hfill
    \begin{subfigure}[b]{0.16\textwidth}
        \begin{quantikz}[row sep={0.7cm,between origins}]
            \lstick{$1$}      & \gate{P} & \\
            \lstick{$\vdots$} &          & \\
            \lstick{$n$}      &	\gate{P} & \\
        \end{quantikz}
        \caption{$\overline{P}^{\otimes k}, P \in \curly{X, Z}$}
    \end{subfigure}
    \vspace{0.5cm}

    \begin{subfigure}[b]{0.17\textwidth}
        \begin{quantikz}[row sep={0.7cm,between origins}]
            \lstick{$1$} &	    & \\
            \lstick{$i+1$} & \gate{P} & \\
            \lstick{$j+1$} & \gate{P} & \\
            \lstick{$n$} &		& \\
        \end{quantikz}
        \caption{$\overline{P}_{ij}, P\in \curly{X, Y, Z}$}
    \end{subfigure}
    \hfill
    \begin{subfigure}[b]{0.29\textwidth}
        \begin{quantikz}[row sep={0.7cm,between origins}]
            \lstick{$1$} &		&	   &	      & \\
            \lstick{$i+1$} & \ctrl{1} & \ctrl{2} &	  & \\
            \lstick{$j+1$} & \phase{} &	       & \ctrl{1} & \\
            \lstick{$n$} & \gate{Z} & \phase{} & \phase{} &\\
        \end{quantikz}
        \caption{$\overline{CZ}_{ij}$}
    \end{subfigure}
    \hfill
    \begin{subfigure}[b]{0.32\textwidth}
        \begin{quantikz}[row sep={0.7cm,between origins}]
            \lstick{$1$}  &	\targ{}   & \targ{}	  & 	      &          & \\
            \lstick{$i+1$}& \ctrl{-1} &           & 	      & \ctrl{1} & \\
            \lstick{$j+1$}&           &	          & \targ{}   & \targ{}  & \\
            \lstick{$n$}  &           & \ctrl{-3} & \ctrl{-1} &          & \\
        \end{quantikz}
        \caption{$\overline{CNOT}_{ij}$}
    \end{subfigure}

    \vspace{0.75cm}
    \begin{subfigure}[b]{0.3\textwidth}
        \begin{quantikz}[row sep={0.7cm,between origins}]
            \lstick{$1$}&\gate[4]{H^{\otimes n}} &\swap{3} & \\
              \lstick{$2$}& & & \\
              \lstick{$\vdots$}& & & \\
            \lstick{$n$}&  & \targX{} &
        \end{quantikz}
        \caption{$\overline{H}^{\otimes k}$}
        \label{fig:synth_had_transversal}
    \end{subfigure}
    \hfill
    \begin{subfigure}[b]{0.6\textwidth}
        \begin{quantikz}[row sep={0.7cm,between origins}]
            \lstick{$1$} &	   & \gate{H} & \ctrl{1} &		&
            \ctrl{1} & \gate{H} &	       & \gate{X} & \\
            \lstick{$i+1$} & \ctrl{1}& \gate{H} & \phase{} & \gate{H} & \phase{}
            & \gate{H} & \ctrl{1} &	 & \\
            \lstick{$n$} & \phase{}&	      & 	 &	    &
            &	   & \phase{} & \gate{Z} &\\
        \end{quantikz}
        \caption{$\overline{H}_i$}
    \end{subfigure}
    \hfill
    \caption{\textbf{Logical Gates}: Logical gate implementations on the $\qcode{n}{n-2}{2}$ code. The figure illustrates the synthesis of various logical gates, including single-qubit gates ($X$, $Y$, $Z$, $S$, $H$), encoded multi-qubit gates (CNOT, CZ). Each subfigure shows the physical circuit for the corresponding logical operation on $n$ physical qubits.}
    \label{fig:gate_compilation}
\end{figure*}
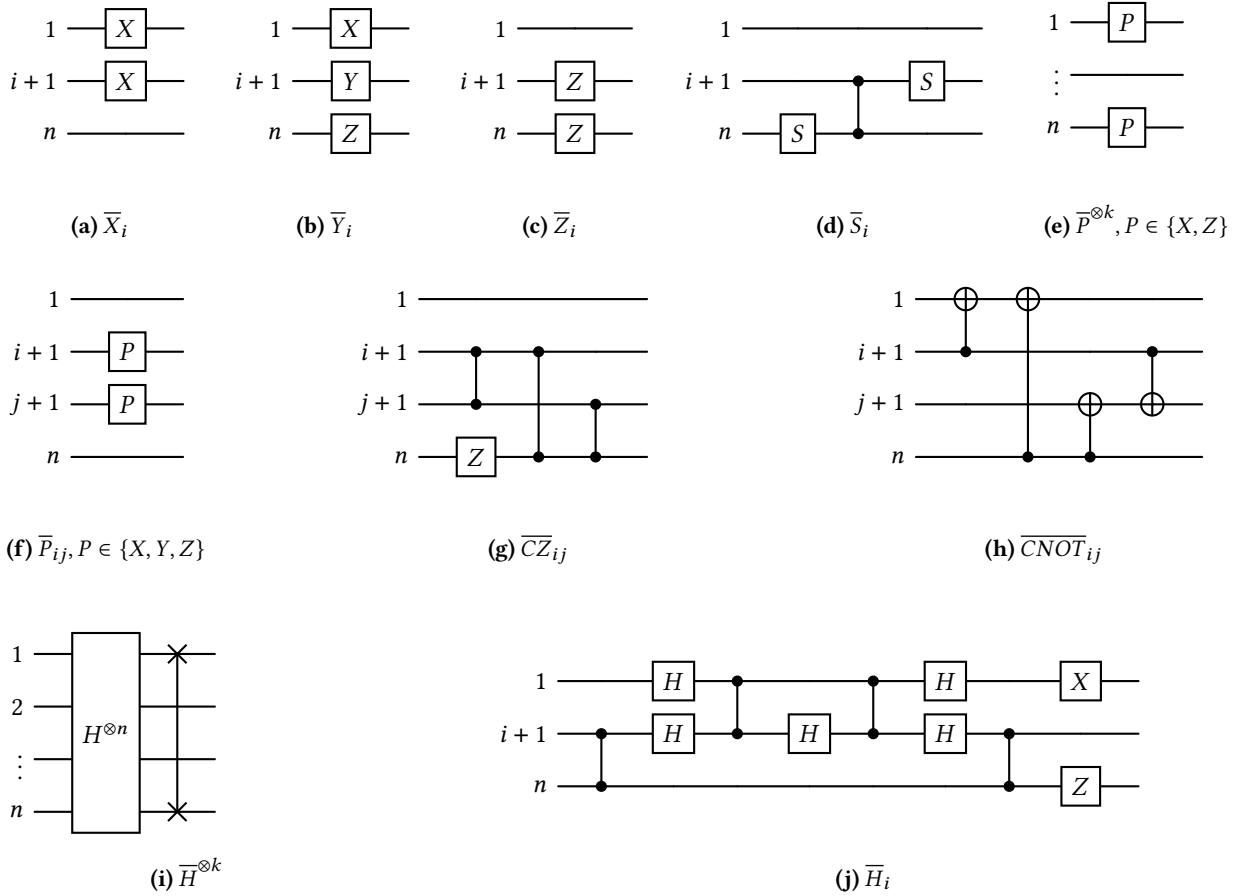

The process is only nearly fault-tolerant as $\text{RZZ}$ gates under the circuit level noise model can result in two- 
qubit correlated errors that are not detectable by the code. However, the motivation for using this construction is 
that (1) $\text{RZZ}$ is native to H2 and other ion traps with high-fidelity, and (2) a fully fault-tolerant protocol 
would likely incur additional overhead losing the already small fault-tolerant abilities of the code. In addition to 
the compilation recipe described above, several other logical gates can be synthesized fault or nearly fault-tolerantly 
as well as more efficiently and/or using alternative native gates. An overview of the constructions used is shown in 
\autoref{fig:gate_compilation}.

\subsection{\label{subsec:mcx}Multi-Controlled Toffoli Gate Synthesis}\label{subsec:label{subsec:mcx}multi-controlled-toffoli-gate-synthesis}

Despite their prevalent use in quantum subroutines including state marking and diffusion in Grover's search, as well as 
more general use cases in quantum RAM and state preparation, synthesizing multi-controlled Toffoli or $C^{\otimes k}X$ 
gates using an elementary gate set (e.g. Clifford+T or CX+U) remains an active area of research
~\cite{algassertConstructingLarge, PhysRevA.106.042602, Satoh_2020, nie2024quantum, orts2022studying, classiqCompetitionSolutions}.

In our circuit compilations, we use the linear depth decomposition provided by~\cite{PhysRevA.106.042602}, which is 
available via \texttt{qclib}~\cite{githubGitHubQclibqclib}. We compute empirically that this decomposition provides a 
1-qubit gate count for $U$ gates as
$
	(\#_U = 16k - 18)
$
2-qubit $CX$ gate count as
$
	(\#_{CX} = 16k - 24)
$
and circuit depth
$
	(D = 32k - 83)
$.
Additionally, the decomposition requires a single \textit{borrowed} ancilla (see \cref{sec:multi-controlled-toffoli-gate-implementations}).

A method proposed by~\cite{Claudon_2024} introduces several families of polylogarithmic-depth $C^{\otimes k}X$ gates, 
and they provide an open-source implementation with an asymptotic depth of $\Theta(\log^3 n)$, also using a single 
\textit{borrowed} ancilla. Despite the asymptotic advantages of~\cite{Claudon_2024}, numerical results show large 
increases in 2-qubit gate count compared to~\cite{PhysRevA.106.042602}, for example, $24.28\times$ at $k = 2^{10}$ and 
$30.58\times$ at $k = 2^{12}$ while circuit depth also remains larger until $k\approx 2^{9}$ which is a far larger 
gate size than that will be used on near-term devices. The full data for this experiment is shown in 
\cref{sec:multi-controlled-toffoli-gate-implementations}.

Another recent approach for MCX decomposition was introduced in~\cite{Nie_2024} and further generalized in
~\cite{khattar_2024}. This method leverages \textit{conditionally clean ancillae} to reduce both gate count and depth 
for various multi-controlled gates. Notably,~\cite{khattar_2024} presents a decomposition of the $C^{\otimes k}X$ gate 
into $2k - 3$ and $4k - 8$ Toffoli gates, achieving an asymptotic depth of $\Theta(\log n)$ with two clean and two dirty
ancillae, respectively. However, in the scope of this work, we use the linear depth decomposition discussed 
~\cite{PhysRevA.106.042602} for its simplicity in FT compilation. 

Compiling the linear depth $C^{\otimes k}X$ gate using the universal procedure outlined in \cref{subsec:compilation} 
gives a compiled gate with CX count $\overline{\#}_{CX}$ and depth $\overline{D}$ as a function of the number of controls $k$:
\begin{align}
	 & \overline{\#}_{CX} = 32k - 44 - (-1)^k                                   \\
	 & \overline{D} = \frac{1}{4} \left( -5 \times (31 + (-1)^k) + 218k \right)
	\label{eq:main_equation}
\end{align}

Asymptotic ratios of the associated fault-tolerant overheads on CX count and depth are, respectively:
\begin{align}
	\pi^\infty_{CX} = 2, \quad \pi^\infty_D = 1.703125
\end{align}

\subsection{Compiling Grover's Algorithm}\label{subsec:compiling-grover's-algorithm}
Grover's algorithm, or quantum search~\cite{grover1996fast}, is a foundational quantum algorithm designed to increase 
the probability of measuring a desired configuration from a set $\mcl{M}$ of \emph{marked} qubit states 
$\ket{\mbf{x}} = \ket{x_0, \cdots, x_n}$, identified by a Boolean oracle $f(\mbf{x}) = \delta_{\mbf{x} \in \mcl{M}}$.
This algorithm iteratively amplifies the probability amplitude of the marked states by applying the Grover operator, 
which consists of the oracle and an inversion about the mean, providing a quadratic speedup over classical search 
methods. Grover's approach extends naturally to amplitude amplification~\cite{Brassard_2002}, a generalization that 
finds broad applications as a subroutine in quantum algorithms.

Building on the compilation methods outlined in \cref{subsec:compilation} and \cref{subsec:mcx}, we detail in 
\cref{tab:grover-comp} the exact resource requirements for the fault-tolerant implementation of Grover's search using 
the $\qcode{n}{n-2}{2}$ code (see \cref{sec:grovers-algorithm} for further details on Grover's algorithm). We list 
resource requirements in terms of the number of qubits, measurements, depth, and universal gate set of the code 
(U and RZZ) which maps closely to the native gate set of today's Ion trap machines (e.g. Quantinuum H2 series
~\cite{quanbtiuumH2}). Additionally, we provide values of the asymptotic overheads introduced by the fault-tolerant 
compilation. These overheads are defined as the ratio of resource usage in the compiled versus uncompiled circuits, in 
the limit where the number of logical qubits $k \to \infty$, and the number of Grover iterations is set to the optimal 
value, \(d = \lfloor(\nicefrac{\pi}{4} \times 2^{k/2} - \nicefrac{1}{2})\rfloor\).

\begin{table*}[htbp]
	\centering
	\setlength{\tabcolsep}{16pt} 
	\renewcommand{\arraystretch}{1.5} 
	\caption{Compiled Circuit resource counts. In addition to the resources specified, each syndrome measurement on the
     compiled circuit requires $4\lceil (k+1)/2 \rceil + 4$ two-qubit gates and 2 measure+reset operations.}
	\label{tab:grover-comp}

	\begin{tabular}{lcc}
		\toprule
		\textbf{Resource}              & \textbf{Compiled Value}                                      & \textbf{Overhead} $(k \to \infty)$ \\
		\midrule
		\multirow{1}{*}{Qubits}        & $2\lceil (k+1)/2 \rceil + 4$                                 & $1$                                \\

		\multirow{1}{*}{Measurements}  & $2\lceil (k+1)/2 \rceil + 3$                                 & $1$                                \\

		\multirow{1}{*}{RZZ Count}     & $10 + (-1)^k - 288d + 2k + 128dk$                            & $4$                                \\

		\multirow{1}{*}{U Count}       & $\frac{1}{2} \left(51 + 5(-1)^k - 228d + 10k + 132dk\right)$ & $1.03125$                          \\

		\multirow{1}{*}{Circuit Depth} & $18 + 2(-1)^k - 394d + 4k + 177dk$                           & $2.765625$                         \\
		\midrule
	\end{tabular}

\end{table*}

\subsection{Optimal Syndrome Scheduling}\label{subsec:optimal-syndrome-scheduling}
Mid-circuit syndrome measurements are an essential resource for the success of quantum error correction as guaranteed 
by the threshold theorem. Syndrome measurements, however, often come at a significant resource demand and hence must be 
appropriately scheduled within a circuit to not add more noise than they can correct for. Furthermore, in the error 
detection regime where post-selection is used to filter out samples \textit{flagged} for error, increasing the frequency
of syndrome extraction boosts the true positive rate for flagging true errors, while also raising the false positive 
rate that comes from flagging errors introduced by the syndrome extraction circuit itself. This issue becomes 
particularly significant when executing a quantum circuit on noisy hardware or with a limited number of shots, as it 
heightens the likelihood that the error detection procedure will filter out all shots, leading to a 
\textit{catastrophic failure} of the algorithm as it produces no clean shots for analysis.

\subsubsection{A Statistical Analysis of Syndrome Scheduling}
We investigate the phenomenon of balancing true and false positive rates of error detection under a simplified 
statistical model that we find to predict valuable insights on optimal syndrome scheduling.
\smallbreak\noindent
Suppose in a given quantum circuit:
\begin{itemize}
	\item An error ($E$) may occur with on the code block probability $\epsilon$
	\item The error detecting protocol will detect (or flag) $E$ it with probability $1 - \delta$
	\item $E$ will be undetected (or pass) with probability $\delta$ (e.g., logical error, or additional error in syndrome readout)
	\item If there is no error on the code block, it is incorrectly flagged as an error with probability $\gamma$ 
    (e.g., due to errors in the syndrome circuit itself)
\end{itemize}

\begin{figure}[htb]
	\centering
	\Tree [.$\mathrm{Circuit}$
		[.$\mathrm{Error}\ \epsilon$
				[.${\mathrm{Flag}\ (1-\delta)}\quad$ ]
					[.${\mathrm{Pass}\ (\delta)}\quad$ ]
			]
			[.$\mathrm{No\ Error}\ (1-\epsilon)$
				[.${\mathrm{Pass}\ (1-\gamma)}\quad$ ]
					[.${\mathrm{Flag}\ (\gamma)}$ ]
			]
	]
	\label{fig:tree}
\end{figure}

The number of correct ($\mathcal{C}$), incorrect ($\mathcal{I}$), and flagged ($\mathcal{F}$) samples out of $N$ shots 
can be modeled by the multinomial distribution with parameters $p_{\mathcal{C}} = (1 - \gamma)(1 - \epsilon)$, 
$p_{\mathcal{I}} = \epsilon \delta$, and $p_{\mathcal{F}} = 1 - p_{\mathcal{I}} - p_{\mathcal{C}} = \gamma + \epsilon -
\gamma \epsilon - \delta \epsilon$. The expected rate of observing the correct output from the circuit after 
post-selecting flagged shots is given by the expected value of $\nicefrac{\mathcal{C}}{(\mathcal{C} + \mathcal{I})}$ 
when the number of accepted shots $\mathcal{C} + \mathcal{I} > 0$ and otherwise $0$ in the event of a 
\textit{catastrophic failure}.

\begin{equation}
	\begin{aligned}
		\mathbb{E}\left[\frac{\mathcal{C}}{\mathcal{C} + \mathcal{I}} \ \text{if} \ \mathcal{C} + \mathcal{I} > 0 \ \text{else} \ 0 \right] = & \\
		\frac{(1-\gamma) (1-\epsilon) (1 - (\gamma - \epsilon (\gamma + \delta) + \epsilon)^N)}{\epsilon (\gamma + \delta - 1) - \gamma + 1}  &
	\end{aligned}
	\label{eq:eq-multinomial}
\end{equation}

The trade-off in QED is that as the true positive rate of correct error detection $1 - \delta$ grows (e.g., by 
increasing the number of syndrome rounds), the false positive rate $\gamma$ of incorrect error detection grows as well. 
Drawing on the statistical model we see that to control for the increasing likelihood of catastrophic failure the only 
option is to increase the number of shots $N$ to counter the effects of increasing likelihood of error in deeper 
circuits. From \cref{eq:eq-multinomial} this means we must keep
\begin{equation}(-\epsilon  (\gamma +\delta )+\gamma +\epsilon )^N\label{eq:catastrophic}\end{equation}
constant which will require $N$ to scale as $O(-\nicefrac{1}{\log(\gamma + \epsilon)})$.

\subsubsection{An Example of Optimal Syndrome Scheduling}\label{subsubsec:syndrome_optimal_eg}
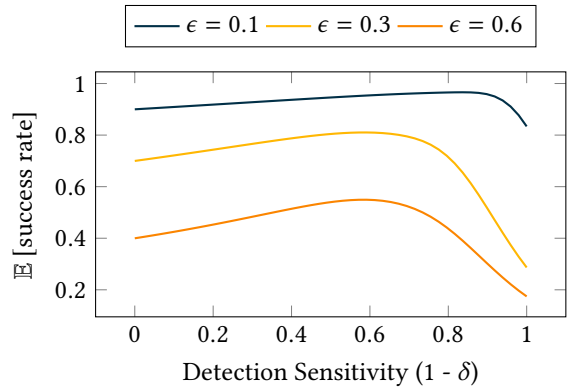
\begin{figure}[!htbp]
	\centering
	\begin{tikzpicture}
		\hspace{-4mm}
		\begin{axis}
			[
				width=\linewidth,
				height=.618\linewidth,
				domain=0:1,
				xlabel={Detection Sensitivity (1 - \(\delta\))},
				ylabel={$\mathbb{E}\left[\text{success rate} \right] $},
				legend style={at={(0.5,1.1)},anchor=south,legend columns=-1},
				grid=none,
				samples=50,
			]

			\addplot[color=custom1, thick]
			{-( (4423.29 * (220265 + 44.8169 * exp(10 * x)) * (-1 + (1e-10 * (4914.77 * x + exp(10 * x) * (9 + x))^10) / (4914.77 + exp(10 * x))^10) ) / ((4914.77 + exp(10 * x)) * (220265 + exp(10 * x) * (4.48169 - 4.48169 * x) - 22026.5 * x)) )};
			\addlegendentry{$\epsilon = 0.1$}

			\addplot[color=custom2, thick]
			{-( (7 * exp(7) * (-1 + (3/10 + 1/(1 + exp(7 - 10 * x)) - 3/10 * (1 + 1/(1 + exp(7 - 10 * x)) - x))^10)) / (exp(7) * (10 - 3 * x) - 3 * exp(10 * x) * (-1 + x)))};
			\addlegendentry{$\epsilon = 0.3$}

			\addplot[color=custom3, thick]
			{-( (2 * exp(7) * (-1 + (3 * exp(7) * x + exp(10 * x) * (2 + 3 * x))^10 / (9765625 * (exp(7) + exp(10 * x))^10))) / (exp(7) * (5 - 3 * x) - 3 * exp(10 * x) * (-1 + x)))};
			\addlegendentry{$\epsilon = 0.6$}

		\end{axis}
	\end{tikzpicture}
	\caption{\textbf{Optimal syndrome scheduling in error-detected circuits.} When a user has the ability to increase 
    the sensitivity of their fault-tolerant scheme for detecting errors through the implementation of syndrome 
    measurements, they will simultaneously increase the false detection of errors via the syndrome circuits themselves. 
    Such a trade-off presents non-obvious optimal syndrome scheduling strategies that will often necessitate experimental
    validation.}
	\label{fig:multinomial}
\end{figure}

To demonstrate how Equation~\ref{eq:eq-multinomial} can generate similar behavior to real experiments that will follow, 
we replace $\gamma$ with a strictly decreasing function of $\delta$ to set the true and false positive rates proportional
to each other:
\begin{equation}
	\gamma \coloneqq \frac{1}{1 + e^{(10 \delta - 3)}}
	\label{eq:gamma-to-delta}
\end{equation}
Under this model, we examine the relationship between the detection sensitivity $1-\delta$ and intrinsic error rate $\epsilon$
with the expected success rate in \cref{eq:eq-multinomial}. \autoref{fig:multinomial} shows that for different values of
$\epsilon$ corresponding to intrinsic hardware noise, there exists an optimal choice of the error detection sensitivity 
$1 - \delta$ to maximize the success rate of the algorithm. This insight reveals that syndrome scheduling---which is the
algorithm designer's tool for controlling the true and false positive rates---is of vital importance for squeezing the 
most out of today's noisy quantum processors. A direct experimental realization of \cref{fig:multinomial} using 
\cref{eq:gamma-to-delta} is shown in \cref{sec:syndrome-experiments}.

\section{Evaluation}\label{sec:evaluation}
\begin{figure*}[!htb]
	\includegraphics[width=\linewidth]{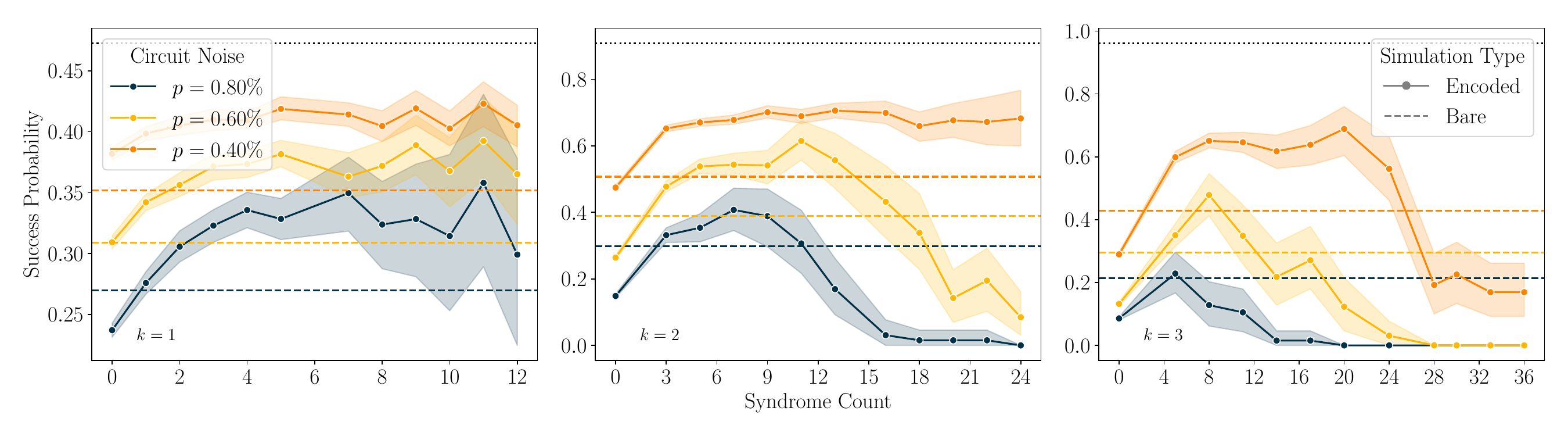}
	\caption{Probability of successful state identification in 4 qubit Grover's search encoded with a \(\qcode{6}{4}{2}\) code under various noise conditions
		$p \in \{0.4\%, 0.6\%, 0.8\% \}$. The simulations delineate the success probabilities across $k = 1, 2, 3$ Grover iterations with different syndrome measurement schedules.
		The dashed lines represent the performance of an unencoded circuit under identical conditions. The uppermost black dashed line indicates the theoretical maximum success probability
		achievable in an ideal, noiseless Grover's search. The shaded areas around each line indicate the 95\% confidence intervals, computed via percentile bootstrap over 50 independent trials.}
	\label{fig:grover_4q}
\end{figure*}

\begin{table*}[!htbp]
	\caption{\textbf{Aggregate performance metrics across noise regimes.} The table compares encoded ($\text{enc}$) and 
    unencoded ($\text{bare}$) success probabilities $\eta$ as well as the normalized improvement metric $\nu$, computed at different syndrome rate optimality levels and grouped into six equally spaced noise bins (values scaled by $10^3$ for readability). 
	$\eta$ is the empirical success probability normalized by the theoretical maximum (or noiseless) probability for that experiment based on the search space dimension and number of marker diffusion iterations (see \cref{sec:grovers-algorithm}). Symbols $\star$, $75$ and $\overline{\cdot}$ indicate that the values are computed respectively at the optimal, $75^{\text{th}}$ percentile, and mean across all syndrome rates tested.
	$\nu$ is a direct measure of the expected improvement from applying fault tolerance, as plotted in \cref{fig:combined-heatmaps} (b). Each metric is described in detail in \cref{subsec:circuit-size-and-quantum-error-detection-performance}. Key metrics are additionally visualized in the subfigure below with the x-axis corresponding to the six noise bins reported.
	Metrics are reported as mean $\pm$ standard error of the mean (SEM). Key trends include: (1) All performance metrics degrade monotonically with increasing noise, (2) normalized improvements $\nu$ become negative at higher noise levels, indicating error correction introduces net overhead, and (3) optimal encoded performance exceeds mean encoded performance by an average of $6.7\times$ across regimes and tends to increase with the noise level. The rightmost column shows population averages across all noise levels tested.}
	\label{tab:qec_results}
	\centering
	\begin{tabular*}{\textwidth}{@{\extracolsep{\fill}}l*{7}{c}@{}}
	\thead{Noise\\Range $10^{-3}$} & \thead{0.25--1.25} & \thead{1.50--2.50} & \thead{2.75--4.00} & \thead{4.25--5.25} & \thead{5.50--6.50} & \thead{6.75--7.75} & \thead{Average} \\
	\midrule[1.5pt]
	\thead{${\eta}^{\star}_\text{enc}$} & 0.875 ± 0.02 & 0.555 ± 0.04 & 0.414 ± 0.04 & 0.309 ± 0.04 & 0.211 ± 0.03 & 0.180 ± 0.02 & 0.417 ± 0.02 \\
	\thead{${\eta}^{75\%}_\text{enc}$} & 0.803 ± 0.02 & 0.417 ± 0.03 & 0.270 ± 0.03 & 0.176 ± 0.03 & 0.113 ± 0.02 & 0.077 ± 0.02 & 0.303 ± 0.01 \\
	\thead{$\overline{\eta}_\text{enc}$} & 0.773 ± 0.03 & 0.372 ± 0.03 & 0.227 ± 0.03 & 0.142 ± 0.02 & 0.093 ± 0.02 & 0.068 ± 0.01 & 0.273 ± 0.01 \\
	\thead{$\overline{\eta}_\text{bare}$} & 0.651 ± 0.03 & 0.326 ± 0.03 & 0.243 ± 0.02 & 0.193 ± 0.02 & 0.166 ± 0.02 & 0.145 ± 0.02 & 0.283 ± 0.01 \\
	\thead{$\nicefrac{\eta^\star_\text{enc}}{\overline{\eta}_\text{bare}}$} & 1.515 ± 0.16 & 4.980 ± 0.43 & 7.131 ± 0.47 & 8.307 ± 0.49 & 8.847 ± 0.47 & 9.263 ± 0.39 & 6.731 ± 0.20 \\\midrule[.5pt]
	\thead{$\nu^{\star}$} & 0.781 ± 0.04 & 0.569 ± 0.08 & 0.330 ± 0.05 & 0.194 ± 0.04 & 0.060 ± 0.02 & 0.042 ± 0.02 & 0.326 ± 0.02 \\
	\thead{$\nu^{75\%}$} & 0.606 ± 0.03 & 0.230 ± 0.03 & 0.039 ± 0.04 & $-$0.061 ± 0.05 & $-$0.122 ± 0.05 & $-$0.129 ± 0.04 & 0.090 ± 0.02 \\
	\thead{$\overline{\nu}$} & 0.489 ± 0.04 & 0.084 ± 0.05 & $-$0.115 ± 0.07 & $-$0.136 ± 0.06 & $-$0.151 ± 0.05 & $-$0.142 ± 0.04 & 0.001 ± 0.02 \\
	\bottomrule
	\end{tabular*}
	\includegraphics[width=\linewidth]{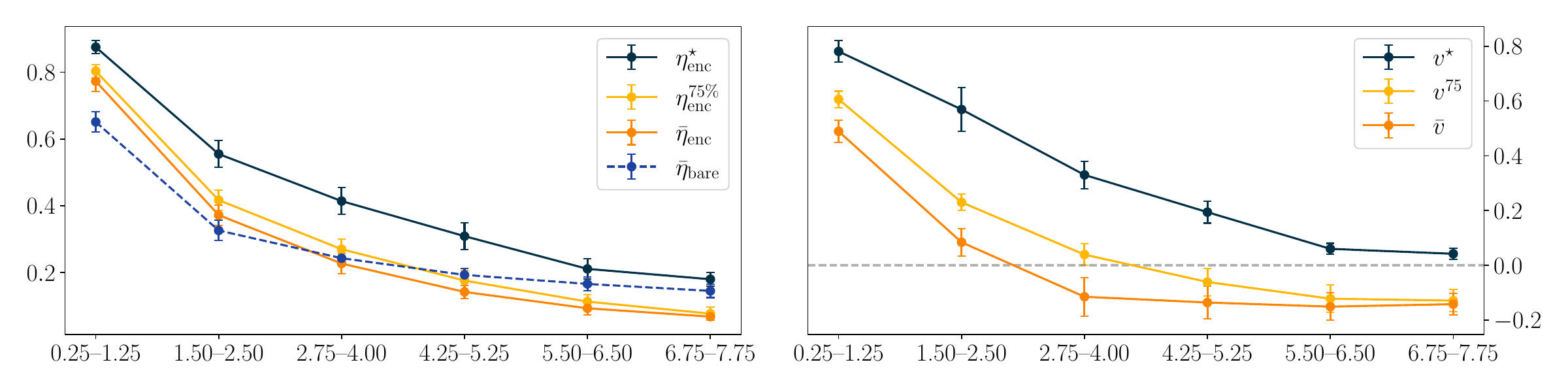}
\end{table*}
We evaluate the performance of $\qcode{n}{n-2}{2}$ codes under a standard circuit-level noise model. In this model, 
state preparation, measurement, single-qubit gates, and two-qubit gates are subject to errors. Each single-qubit gate 
is followed by a Pauli error $\{X, Y, Z\}$ occurring with probability $p/3$, while two-qubit gates incur correlated 
Pauli errors from $\{I, X, Y, Z\}^{\otimes 2} \setminus \{I \otimes I\}$ with probability $p/15$. State preparation and 
measurement (SPAM) errors are modeled as bit-flips with probability $p$. This framework captures platforms where gate 
and measurement errors dominate over idling noise.

Our experiments focus on Grover's algorithm, implemented on the $s$-qubit search space with an oracle marking the 
state $\ket{\mathbf{x}} = \ket{1}^{\otimes s}$ via a $C^{s-1}Z$ gate. This choice simplifies circuit structure while 
reflecting worst-case success probabilities. For the $\qcode{n}{n-2}{2}$ code, logical Hadamard gates are realized 
transversally (see \autoref{fig:synth_had_transversal}), minimizing overhead in the diffusion operator. Toffoli gates 
are decomposed using the linear-depth method from~\cite{PhysRevA.106.042602}, requiring one borrowed ancilla.

As the code requires an even number of logical qubits we must round up odd-sized circuits (after adding the ancilla). We
also require two more ancillae for flag measurements and syndrome readout, hence our implementation of Grover's 
algorithm demands
\begin{equation}  
\text{physical qubits} = s + \frac{1}{2} \left((-1)^{s} + 7\right)  
\end{equation}  
For each circuit and noise level $p$ in our experiments that follow, we perform 50 trials of 1,000 shots each. We 
construct 95\% confidence intervals in figures via percentile bootstrap. Post-selection is applied to discard runs 
flagged by syndrome measurements at all steps.

\subsection{Syndrome Scheduling}\label{subsec:syndrome-scheduling}

We begin with an empirical identification of optimal syndrome measurement frequencies in a simplified setting. We 
evaluate Grover's search encoded with a $\qcode{6}{4}{2}$ code under a circuit-level noise model with error rates 
$p = 0.4\%, 0.6\%, \text{ and } 0.8\%$. We vary the circuit depth by tuning the number of Grover iterations 
$k \in \{1, 2, 3\}$.  

As shown in \cref{fig:grover_4q}, increasing the syndrome measurement frequency initially improves success probabilities
by suppressing error propagation. However, beyond a critical point, additional syndrome rounds introduce diminishing 
returns due to overhead. For example, at $p = 0.6\%$ and $k=2$, success probabilities peak at 12 syndrome rounds before 
declining. 

The encoded circuits consistently outperform unencoded baselines (dashed lines) in low-to-moderate noise regimes, 
often achieving 2$\times$ higher success rates, especially for shallow circuits ($k=1$). However, deeper circuits 
($k=3$) exhibit narrower advantage windows, as accumulated noise erodes the benefits of error detection. These results 
demonstrate that optimal syndrome scheduling is context-dependent, as investigated by our statistical results in 
\cref{subsec:optimal-syndrome-scheduling}, requiring joint optimization with noise levels and algorithm depth to maximize 
fault-tolerant advantage.

\subsubsection{Mitigating Catastrophic Failure}
\begin{figure}[!htb]
	\includegraphics[width=\linewidth]{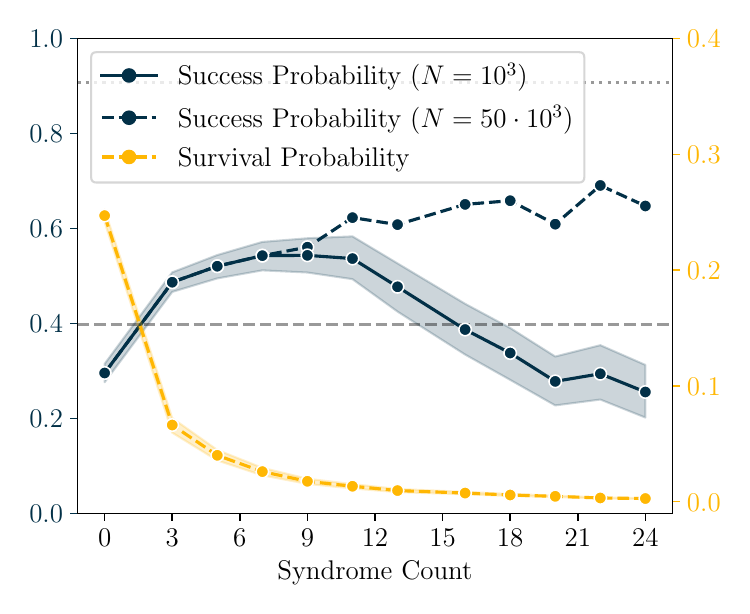}
	\caption{\textbf{Shot count on success probability.} We show $k=2$ iteration Grover search on the \(\qcode{6}{4}{2}\)
    code at noise levels averaged over runs with $p \in \{0.4\%, 0.6\%, 0.8\%\}$ (see the setup in \cref{fig:grover_4q}). The
    success probability is shown with both $10^3$ shots and $50\times 10^3$ shots per round to emphasize that in 
    shot-limited environments adding syndrome measurements yields diminishing returns compared to the "shot-rich 
    environment"---an effect explained in this work by the phenomenon of \textit{catastrophic failures}. Survival probability
    (the fraction of shots that make it past post-selection) for the $50\times 10^3$ shot run is shown on the 
    complementary axis. The dashed grey line shows the average success rate of the unencoded simulation under the same 
    noise conditions.}
	\label{fig:catastrophic}
\end{figure}

Catastrophic failure, which we recall to be the case where an experiment produces no usable shots, is the main factor 
that limits the scaling of simply applying more syndrome measurements. In \cref{fig:catastrophic} we experiment with 
the relationship between shot count and scalability of syndrome extraction. A 50-fold increase in the number of shots 
per experiment allows syndrome measurements to scale to considerably higher success rates despite a rapidly decreasing 
survival rate of non-flagged shots. For higher noise levels and deeper circuits, this will not always be practical as the
number of shots will need to scale asymptotically as $O(\nicefrac{-1}{\log(2p)})$ (\cref{eq:catastrophic}) to keep the 
probability of catastrophic failure bounded.

\begin{figure*}
	\centering
	\includegraphics[width=\textwidth]{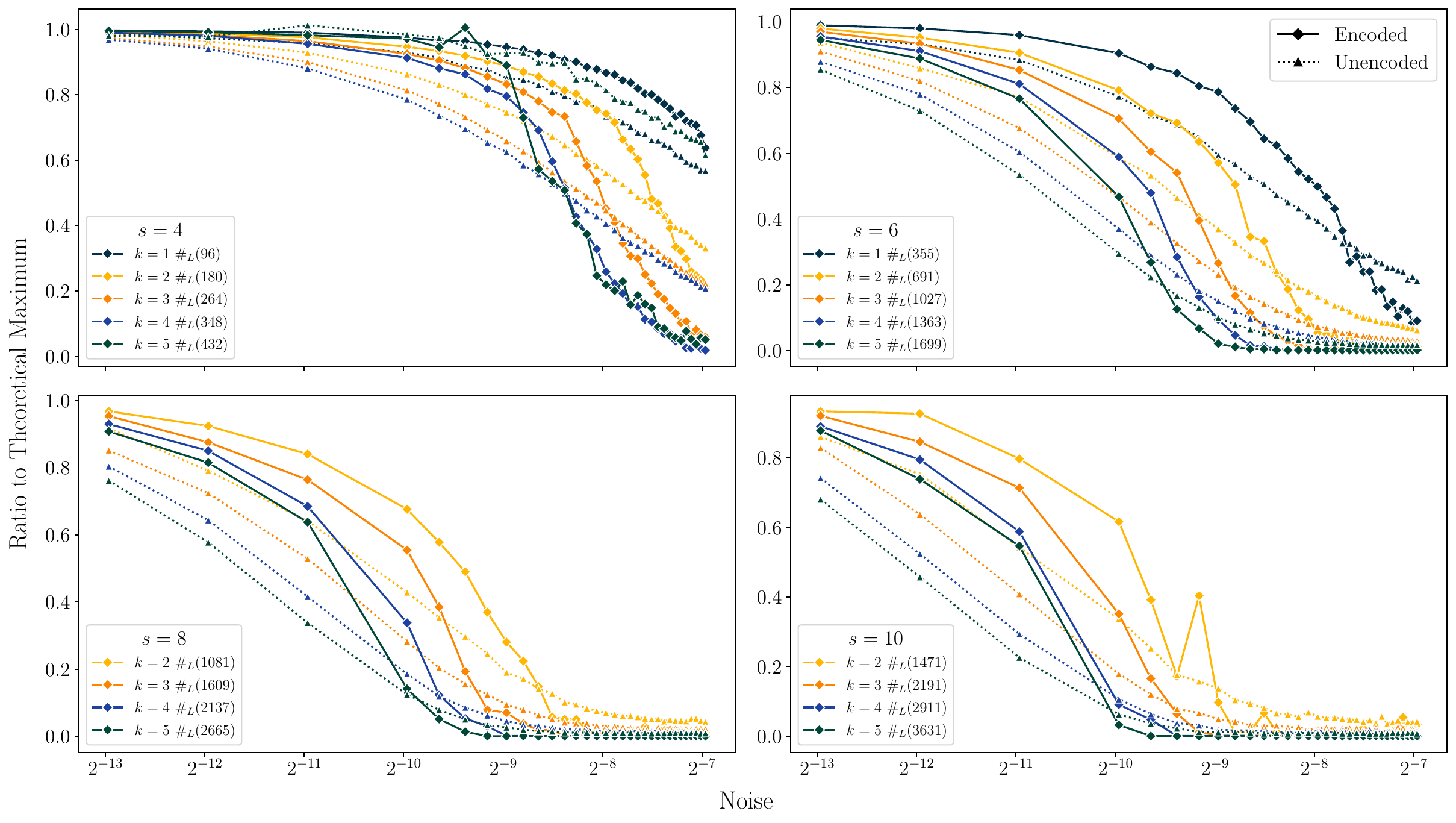}
	\caption{Ratio of success probability to the theoretical maximum for encoded ($\eta_{\text{enc}}$) and unencoded
    ($\eta_{\text{unenc}}$) Grover circuits across varying noise levels. Solid lines with diamonds represent encoded
    circuits, while dotted lines with triangles denote unencoded circuits. We omit $k=1$ for the Grover search spaces
    $s = 8$ and $s=10$ due to near-zero ideal success probabilities, which would otherwise cause large spikes in
    the ratios.}
	\label{fig:ratio_th_max}
\end{figure*}

\subsection{Circuit Size and Performance}
\label{subsec:circuit-size-and-quantum-error-detection-performance}

We now present our main experimental results, focusing on how \emph{circuit size} (the total count of gates, measurements,
and state preparations) affects the performance of quantum error detection (QED). Specifically, we examine the interplay 
between noise level, syndrome measurement frequency, and algorithm depth in Grover's search, using the 
\(\qcode{n}{n-2}{2}\) code under a circuit-level noise model. 

\paragraph{Experimental Setup.}
In this set of experiments we vary the circuit-level noise parameter \(p\) from \(0.0125\%\) to \(0.8\%\) in steps of $0.025\%$. For search space dimension \(s \in \{4,6,8,10\}\) and Grover iteration count \(k \in \{1,\dots,5\}\), we sweep syndrome measurement schedules from 1 to 12 logical operations. All flagged samples from any syndrome round are discarded before analysis. This setup enables us to thoroughly examine how noise, circuit depth, and syndrome rate jointly influence the performance of quantum error detection. 

\paragraph{Reported Metrics.} 
In addition to the raw success probabilities, we use two primary metrics to quantify the efficacy of QED. First, we define 
\(\eta\), the ratio of a circuit's observed success probability to the \emph{theoretical maximum} (i.e., the noiseless ideal) 
for the same circuit parameters:
\[
  \eta_{\text{enc}} \;=\; \frac{p_{\text{enc}}}{p_{\text{ideal}}}
  \quad\text{and}\quad
  \eta_{\text{bare}} \;=\; \frac{p_{\text{bare}}}{p_{\text{ideal}}}.
\]
Here, \(p_{\text{enc}}\) and \(p_{\text{bare}}\) are the measured success probabilities of the encoded and unencoded 
implementations, respectively, and \(p_{\text{ideal}}\) is the success probability when all operations are noiseless.

Second, to highlight the relative improvement of QED, we introduce
\[
  \nu \;=\; \frac{\,p_{\text{enc}} \;-\; p_{\text{bare}}\,}{\,p_{\text{ideal}} \;-\; p_{\text{bare}}\,}.
\]
This \emph{normalized improvement} metric measures how much the encoded circuit's success probability is boosted compared 
to the gap between unencoded and ideal performance. When \(\nu>0\), QED confers a net advantage; negative \(\nu\) values 
indicate that overhead from error detection outweighs its benefits.

In the reported metrics in \cref{tab:qec_results} we include subscripts $\star$, $\overline{\cdot}$ and $75\%$ to indicate metrics calculated at the \textit{optimal}, \textit{mean}, and $75^{\text{th}}$ percentile over the range of syndrome schedules in our experiments. 

\paragraph{Overview of Results.}
\cref{fig:ratio_th_max} shows the empirical ratios \(\eta\) for various circuit sizes (i.e., different search space dimensions and 
numbers of Grover iterations) across a range of noise values. Consistent with the heatmaps in 
\cref{fig:combined-heatmaps} and the aggregate data in \cref{tab:qec_results}, we find a clear region in which 
the encoded circuits (\(\eta_{\text{enc}}\)) exceed their unencoded counterparts (\(\eta_{\text{bare}}\)), confirming that 
QED can substantially suppress errors when noise is moderate, circuits are shallow, and syndrome measurements are carefully scheduled. 

However, as the noise grows, the benefit from error detection ultimately saturates and then diminishes. Beyond a certain 
circuit-dependent \emph{critical noise threshold}, overheads from frequent syndrome checks begin to dominate, and success
probabilities for both encoded and bare circuits collapse toward zero. These observations are reflected in the \(\nu\)
values reported in \cref{tab:qec_results}, where \(\nu\) eventually becomes negative for sufficiently large noise. 

Our main results are also summarized visually in \cref{fig:combined-heatmaps}, which shows the region of positive
$\nu^{\star}$ explicitly across the entire domain of our experimental parameters (noise levels and circuit size).
To extend our experimental findings over a continuous range of parameters, we fit our results using a four-layer neural network with hidden layer dimension 256.
This interpolation allows us to estimate optimal syndrome rates and expected success rate improvements directly as functions of the experimental parameters.
While this set of experiments is governed by only two parameters (circuit size and noise), our approach could easily generalize to predicting optimal fault-tolerant
compilation parameters and expected performance metrics for higher-dimensional feature spaces. Future work will explore extending this methodology to
more complex experimental domains, enabling data-driven optimization of fault-tolerant quantum protocols.  

\paragraph{Scaling and Limitations.}
An important takeaway from our simulations is that the success of QED depends sensitively on both the total number of 
operations and the available \emph{shot budget}. As circuit size increases, more syndrome extractions are typically 
required to keep error accumulation in check, but each extraction introduces its own error risk and reduces the fraction 
of surviving (unflagged) shots. \cref{fig:combined-heatmaps}(a) illustrates how the \emph{optimal} number of 
syndrome measurements varies with circuit depth and noise, while \cref{fig:combined-heatmaps}(b) highlights the 
associated improvements in success probability $\nu$ when the optimal syndrome schedule is used. Together, these results
underscore that \emph{syndrome scheduling} must be carefully tailored to hardware noise characteristics and experiment
constraints to reap the advantages of QED.

Further details on syndrome scheduling, shot budgets, and the interplay of these factors appear in 
\cref{sec:experiment-results}.

\section{Conclusion}\label{sec:conclusion}
In this work, we have demonstrated the utility of quantum error detection (QED) using $\qcode{n}{n-2}{2}$ codes as a 
resource-efficient strategy for early fault-tolerant quantum experiments. By developing an end-to-end fault-tolerant 
compilation pipeline and simulating Grover's algorithm under a circuit-level noise model, we empirically identified 
non-trivial regimes where QED provides measurable improvements in success probabilities. Our results primarily highlight
the important role of syndrome scheduling when balancing error detection efficacy against the overhead of increased 
operations and mid-circuit measurements to mitigate catastrophic failure. Encoded circuits, when tuned to noise levels 
and shot budgets, consistently outperform unencoded implementations, achieving up to a $2\times$ improvement in success 
rates for shallow circuits.  

While QED alone cannot scale to commercially relevant problem sizes due to clear statistical limitations, it may still
offer a pragmatic advantage as a \emph{constant-overhead error mitigation layer}. For instance, a single concatenation 
of QED onto existing fault-tolerant protocols could enhance noise resilience—effectively acting as logical quantum error
detection to suppress uncorrected errors in larger fault-tolerant demonstrations. This approach aligns with the growing 
emphasis on error correction strategies tailored to hardware constraints.  

The experiments presented in this work are notably limited by the assumptions of the circuit-level noise model and hence
real-world performance will vary due to hardware-specific noise channels. Our primary goal in this work was to study 
the compilation strategies, experimental procedures, and metrics--such as optimal syndrome rate calculations and 
fault-tolerant improvement ratios—that are directly transferable to experiments on real hardware given the resources. 
To enable community validation and extension, we open-source our experimental data, processing pipeline and release 
an initial version of a $\qcode{n}{n-2}{2}$ code logical compiler at \href{https://codeqraft.xyz}{codeqraft.xyz}.

\section{Acknowledgments}
The authors would like to thank Jan Tułowiecki, Ákos Nagy, Hirsh Kamakari, Emil Żak and the BEIT Inc. team for helpful discussions and feedback. 
This work is additionally dedicated to Wojciech Burkot (co-founder of BEIT) who participated in several of the initial discussions for the project and passed away sadly in late 2024.

\bibliographystyle{plainnat}     
\bibliography{main}

\begin{thebibliography}{43}
\providecommand{\natexlab}[1]{#1}
\providecommand{\url}[1]{\texttt{#1}}
\expandafter\ifx\csname urlstyle\endcsname\relax
  \providecommand{\doi}[1]{doi: #1}\else
  \providecommand{\doi}{doi: \begingroup \urlstyle{rm}\Url}\fi

\bibitem[Acharya et~al.(2024)Acharya, Aghababaie-Beni, Aleiner, Andersen,
  Ansmann, Arute, Arya, Asfaw, Astrakhantsev, Atalaya,
  et~al.]{acharya2024quantum}
Rajeev Acharya, Laleh Aghababaie-Beni, Igor Aleiner, Trond~I Andersen, Markus
  Ansmann, Frank Arute, Kunal Arya, Abraham Asfaw, Nikita Astrakhantsev, Juan
  Atalaya, et~al.
\newblock Quantum error correction below the surface code threshold.
\newblock \emph{arXiv preprint arXiv:2408.13687}, 2024.

\bibitem[Anderson et~al.(2014)Anderson, Duclos-Cianci, and
  Poulin]{anderson2014fault}
Jonas~T Anderson, Guillaume Duclos-Cianci, and David Poulin.
\newblock Fault-tolerant conversion between the steane and reed-muller quantum
  codes.
\newblock \emph{Physical review letters}, 113\penalty0 (8):\penalty0 080501,
  2014.

\bibitem[Araujo et~al.(2023)Araujo, Araújo, da~Silva,
  et~al.]{githubGitHubQclibqclib}
Israel~F. Araujo, Ismael C.~S. Araújo, Leon~D. da~Silva, et~al.
\newblock {G}it{H}ub - qclib/qclib, 2023.
\newblock URL \url{https://github.com/qclib/qclib}.

\bibitem[Bluvstein et~al.(2024)Bluvstein, Evered, Geim, Li, Zhou, Manovitz,
  Ebadi, Cain, Kalinowski, Hangleiter, et~al.]{bluvstein2024logical}
Dolev Bluvstein, Simon~J Evered, Alexandra~A Geim, Sophie~H Li, Hengyun Zhou,
  Tom Manovitz, Sepehr Ebadi, Madelyn Cain, Marcin Kalinowski, Dominik
  Hangleiter, et~al.
\newblock Logical quantum processor based on reconfigurable atom arrays.
\newblock \emph{Nature}, 626\penalty0 (7997):\penalty0 58--65, 2024.

\bibitem[Brassard et~al.(2002)Brassard, Høyer, Mosca, and Tapp]{Brassard_2002}
Gilles Brassard, Peter Høyer, Michele Mosca, and Alain Tapp.
\newblock Quantum amplitude amplification and estimation, 2002.
\newblock ISSN 0271-4132.
\newblock URL \url{http://dx.doi.org/10.1090/conm/305/05215}.

\bibitem[Chao and Reichardt(2018)]{flags}
Rui Chao and Ben~W. Reichardt.
\newblock Quantum error correction with only two extra qubits.
\newblock \emph{Physical Review Letters}, 121\penalty0 (5), August 2018.
\newblock ISSN 1079-7114.
\newblock \doi{10.1103/physrevlett.121.050502}.
\newblock URL \url{http://dx.doi.org/10.1103/PhysRevLett.121.050502}.

\bibitem[Chen et~al.(2023)Chen, Nielsen, Ebert, Inlek, Wright, Chaplin,
  Maksymov, P{\'a}ez, Poudel, Maunz, et~al.]{chen2023benchmarking}
Jwo-Sy Chen, Erik Nielsen, Matthew Ebert, Volkan Inlek, Kenneth Wright,
  Vandiver Chaplin, Andrii Maksymov, Eduardo P{\'a}ez, Amrit Poudel, Peter
  Maunz, et~al.
\newblock Benchmarking a trapped-ion quantum computer with 29 algorithmic
  qubits.
\newblock \emph{arXiv preprint arXiv:2308.05071}, 2023.

\bibitem[Chen and Rengaswamy(2024)]{chen2024tailoring}
Zhuangzhuang Chen and Narayanan Rengaswamy.
\newblock Tailoring fault-tolerance to quantum algorithms.
\newblock \emph{arXiv preprint arXiv:2404.11953}, 2024.

\bibitem[Classiq(2022)]{classiqCompetitionSolutions}
Classiq.
\newblock Competition solutions: Decomposing a multi-controlled toffoli gate,
  2022.
\newblock URL \url{https://www.classiq.io/insights/competition-results-mcx}.

\bibitem[Claudon et~al.(2024)Claudon, Zylberman, Feniou, Debbasch, Peruzzo, and
  Piquemal]{Claudon_2024}
Baptiste Claudon, Julien Zylberman, César Feniou, Fabrice Debbasch, Alberto
  Peruzzo, and Jean-Philip Piquemal.
\newblock Polylogarithmic-depth controlled-not gates without ancilla qubits.
\newblock \emph{Nature Communications}, 15\penalty0 (1), July 2024.
\newblock ISSN 2041-1723.
\newblock \doi{10.1038/s41467-024-50065-x}.
\newblock URL \url{http://dx.doi.org/10.1038/s41467-024-50065-x}.

\bibitem[da~Silva and Park(2022)]{PhysRevA.106.042602}
Adenilton~J. da~Silva and Daniel~K. Park.
\newblock Linear-depth quantum circuits for multiqubit controlled gates.
\newblock \emph{Phys. Rev. A}, 106:\penalty0 042602, Oct 2022.
\newblock \doi{10.1103/PhysRevA.106.042602}.
\newblock URL \url{https://link.aps.org/doi/10.1103/PhysRevA.106.042602}.

\bibitem[DeCross et~al.(2024)DeCross, Haghshenas, Liu, Alexeev, Baldwin,
  Bartolotta, Bohn, Chertkov, Colina, DelVento,
  et~al.]{decross2024computational}
Matthew DeCross, Reza Haghshenas, Minzhao Liu, Yuri Alexeev, Charles~H Baldwin,
  John~P Bartolotta, Matthew Bohn, Eli Chertkov, Jonhas Colina, Davide
  DelVento, et~al.
\newblock The computational power of random quantum circuits in arbitrary
  geometries.
\newblock \emph{arXiv preprint arXiv:2406.02501}, 2024.

\bibitem[Dennis et~al.(2002)Dennis, Kitaev, Landahl, and
  Preskill]{dennis2002topological}
Eric Dennis, Alexei Kitaev, Andrew Landahl, and John Preskill.
\newblock Topological quantum memory.
\newblock \emph{Journal of Mathematical Physics}, 43\penalty0 (9):\penalty0
  4452--4505, 2002.

\bibitem[Gidney(2015)]{algassertConstructingLarge}
Craig Gidney.
\newblock {C}onstructing {L}arge {C}ontrolled {N}ots --- algassert.com, 2015.
\newblock URL
  \url{https://algassert.com/circuits/2015/06/05/Constructing-Large-Controlled-Nots.html}.

\bibitem[Gidney et~al.(2024)Gidney, Shutty, and Jones]{gidney2024magic}
Craig Gidney, Noah Shutty, and Cody Jones.
\newblock Magic state cultivation: growing t states as cheap as cnot gates.
\newblock \emph{arXiv preprint arXiv:2409.17595}, 2024.

\bibitem[Goto(2024)]{goto2024manyhypercube}
Hayato Goto.
\newblock Many-hypercube codes: High-rate quantum error-correcting codes for
  high-performance fault-tolerant quantum computation, 2024.

\bibitem[Gottesman(1998)]{gottesman1998theory}
Daniel Gottesman.
\newblock Theory of fault-tolerant quantum computation.
\newblock \emph{Physical Review A}, 57\penalty0 (1):\penalty0 127, 1998.

\bibitem[Gottesman(2013)]{gottesman2013fault}
Daniel Gottesman.
\newblock Fault-tolerant quantum computation with constant overhead.
\newblock \emph{arXiv preprint arXiv:1310.2984}, 2013.

\bibitem[Grover(1996)]{grover1996fast}
Lov~K. Grover.
\newblock A fast quantum mechanical algorithm for database search, 1996.

\bibitem[Hangleiter et~al.(2024)Hangleiter, Kalinowski, Bluvstein, Cain,
  Maskara, Gao, Kubica, Lukin, and Gullans]{hangleiter2024faulttolerant}
Dominik Hangleiter, Marcin Kalinowski, Dolev Bluvstein, Madelyn Cain, Nishad
  Maskara, Xun Gao, Aleksander Kubica, Mikhail~D. Lukin, and Michael~J.
  Gullans.
\newblock Fault-tolerant compiling of classically hard iqp circuits on
  hypercubes, 2024.

\bibitem[He et~al.(2024)He, Amaro, Shaydulin, and Pistoia]{he2024performance}
Zichang He, David Amaro, Ruslan Shaydulin, and Marco Pistoia.
\newblock Performance of quantum approximate optimization with quantum error
  detection.
\newblock \emph{arXiv preprint arXiv:2409.12104}, 2024.

\bibitem[Hong et~al.(2024)Hong, Durso-Sabina, Hayes, and
  Lucas]{hong2024entangling}
Yifan Hong, Elijah Durso-Sabina, David Hayes, and Andrew Lucas.
\newblock Entangling four logical qubits beyond break-even in a nonlocal code.
\newblock \emph{arXiv preprint arXiv:2406.02666}, 2024.

\bibitem[Horsman et~al.(2012)Horsman, Fowler, Devitt, and
  Van~Meter]{horsman2012surface}
Dominic Horsman, Austin~G Fowler, Simon Devitt, and Rodney Van~Meter.
\newblock Surface code quantum computing by lattice surgery.
\newblock \emph{New Journal of Physics}, 14\penalty0 (12):\penalty0 123011,
  2012.

\bibitem[Javadi-Abhari et~al.(2024)Javadi-Abhari, Treinish, Krsulich, Wood,
  Lishman, Gacon, Martiel, Nation, Bishop, Cross, et~al.]{javadi2024quantum}
Ali Javadi-Abhari, Matthew Treinish, Kevin Krsulich, Christopher~J Wood, Jake
  Lishman, Julien Gacon, Simon Martiel, Paul~D Nation, Lev~S Bishop, Andrew~W
  Cross, et~al.
\newblock Quantum computing with qiskit.
\newblock \emph{arXiv preprint arXiv:2405.08810}, 2024.

\bibitem[Jochym-O’Connor and Laflamme(2014)]{jochym2014using}
Tomas Jochym-O’Connor and Raymond Laflamme.
\newblock Using concatenated quantum codes for universal fault-tolerant quantum
  gates.
\newblock \emph{Physical review letters}, 112\penalty0 (1):\penalty0 010505,
  2014.

\bibitem[Katabarwa et~al.(2024)Katabarwa, Gratsea, Caesura, and
  Johnson]{PRXQuantum.5.020101}
Amara Katabarwa, Katerina Gratsea, Athena Caesura, and Peter~D. Johnson.
\newblock Early fault-tolerant quantum computing.
\newblock \emph{PRX Quantum}, 5:\penalty0 020101, Jun 2024.
\newblock \doi{10.1103/PRXQuantum.5.020101}.
\newblock URL \url{https://link.aps.org/doi/10.1103/PRXQuantum.5.020101}.

\bibitem[Khattar and Gidney(2024)]{khattar_2024}
Tanuj Khattar and Craig Gidney.
\newblock Rise of conditionally clean ancillae for optimizing quantum circuits,
  2024.
\newblock URL \url{https://arxiv.org/abs/2407.17966}.

\bibitem[Linke et~al.(2017)Linke, Gutierrez, Landsman, Figgatt, Debnath, Brown,
  and Monroe]{Linke_2017}
Norbert~M. Linke, Mauricio Gutierrez, Kevin~A. Landsman, Caroline Figgatt,
  Shantanu Debnath, Kenneth~R. Brown, and Christopher Monroe.
\newblock Fault-tolerant quantum error detection.
\newblock \emph{Science Advances}, 3\penalty0 (10), October 2017.
\newblock ISSN 2375-2548.
\newblock \doi{10.1126/sciadv.1701074}.
\newblock URL \url{http://dx.doi.org/10.1126/sciadv.1701074}.

\bibitem[Moses et~al.(2023{\natexlab{a}})Moses, Baldwin, Allman, Ancona,
  Ascarrunz, Barnes, Bartolotta, Bjork, Blanchard, Bohn, et~al.]{moses2023race}
Steven~A Moses, Charles~H Baldwin, Michael~S Allman, R~Ancona, L~Ascarrunz,
  C~Barnes, J~Bartolotta, B~Bjork, P~Blanchard, M~Bohn, et~al.
\newblock A race-track trapped-ion quantum processor.
\newblock \emph{Physical Review X}, 13\penalty0 (4):\penalty0 041052,
  2023{\natexlab{a}}.

\bibitem[Moses et~al.(2023{\natexlab{b}})Moses, Baldwin, Allman, Ancona,
  Ascarrunz, Barnes, Bartolotta, Bjork, Blanchard, Bohn, Bohnet, Brown,
  Burdick, Burton, Campbell, Campora, Carron, Chambers, Chan, Chen,
  Chernoguzov, Chertkov, Colina, Curtis, Daniel, DeCross, Deen, Delaney,
  Dreiling, Ertsgaard, Esposito, Estey, Fabrikant, Figgatt, Foltz, Foss-Feig,
  Francois, Gaebler, Gatterman, Gilbreth, Giles, Glynn, Hall, Hankin, Hansen,
  Hayes, Higashi, Hoffman, Horning, Hout, Jacobs, Johansen, Jones, Karcz,
  Klein, Lauria, Lee, Liefer, Lu, Lucchetti, Lytle, Malm, Matheny, Mathewson,
  Mayer, Miller, Mills, Neyenhuis, Nugent, Olson, Parks, Price, Price, Pugh,
  Ransford, Reed, Roman, Rowe, Ryan-Anderson, Sanders, Sedlacek, Shevchuk,
  Siegfried, Skripka, Spaun, Sprenkle, Stutz, Swallows, Tobey, Tran, Tran,
  Vogt, Volin, Walker, Zolot, and Pino]{Moses_2023}
S. A. Moses, C. H. Baldwin, M. S. Allman, R.~Ancona, L.~Ascarrunz,
  C.~Barnes, J.~Bartolotta, B.~Bjork, P.~Blanchard, M.~Bohn, J. G. Bohnet,
  N. C. Brown, N. Q. Burdick, W. C. Burton, S. L. Campbell, J. P.
  Campora, C.~Carron, J.~Chambers, J. W. Chan, Y. H. Chen, A.~Chernoguzov,
  E.~Chertkov, J.~Colina, J. P. Curtis, R.~Daniel, M.~DeCross, D.~Deen,
  C.~Delaney, J. M. Dreiling, C. T. Ertsgaard, J.~Esposito, B.~Estey,
  M.~Fabrikant, C.~Figgatt, C.~Foltz, M.~Foss-Feig, D.~Francois, J. P.
  Gaebler, T. M. Gatterman, C. N. Gilbreth, J.~Giles, E.~Glynn, A.~Hall,
  A. M. Hankin, A.~Hansen, D.~Hayes, B.~Higashi, I. M. Hoffman, B.~Horning,
  J. J. Hout, R.~Jacobs, J.~Johansen, L.~Jones, J.~Karcz, T.~Klein,
  P.~Lauria, P.~Lee, D.~Liefer, S. T. Lu, D.~Lucchetti, C.~Lytle, A.~Malm,
  M.~Matheny, B.~Mathewson, K.~Mayer, D. B. Miller, M.~Mills, B.~Neyenhuis,
  L.~Nugent, S.~Olson, J.~Parks, G. N. Price, Z.~Price, M.~Pugh, A.~Ransford,
  A. P. Reed, C.~Roman, M.~Rowe, C.~Ryan-Anderson, S.~Sanders, J.~Sedlacek,
  P.~Shevchuk, P.~Siegfried, T.~Skripka, B.~Spaun, R. T. Sprenkle, R. P.
  Stutz, M.~Swallows, R. I. Tobey, A.~Tran, T.~Tran, E.~Vogt, C.~Volin,
  J.~Walker, A. M. Zolot, and J. M. Pino.
\newblock A race-track trapped-ion quantum processor.
\newblock \emph{Physical Review X}, 13\penalty0 (4), December
  2023{\natexlab{b}}.
\newblock ISSN 2160-3308.
\newblock \doi{10.1103/physrevx.13.041052}.
\newblock URL \url{http://dx.doi.org/10.1103/PhysRevX.13.041052}.

\bibitem[Moses et~al.(2023{\natexlab{c}})Moses, Baldwin, Allman, Ancona,
  Ascarrunz, Barnes, Bartolotta, Bjork, Blanchard, Bohn, Bohnet, Brown,
  Burdick, Burton, Campbell, Campora, Carron, Chambers, Chan, Chen,
  Chernoguzov, Chertkov, Colina, Curtis, Daniel, DeCross, Deen, Delaney,
  Dreiling, Ertsgaard, Esposito, Estey, Fabrikant, Figgatt, Foltz, Foss-Feig,
  Francois, Gaebler, Gatterman, Gilbreth, Giles, Glynn, Hall, Hankin, Hansen,
  Hayes, Higashi, Hoffman, Horning, Hout, Jacobs, Johansen, Jones, Karcz,
  Klein, Lauria, Lee, Liefer, Lu, Lucchetti, Lytle, Malm, Matheny, Mathewson,
  Mayer, Miller, Mills, Neyenhuis, Nugent, Olson, Parks, Price, Price, Pugh,
  Ransford, Reed, Roman, Rowe, Ryan-Anderson, Sanders, Sedlacek, Shevchuk,
  Siegfried, Skripka, Spaun, Sprenkle, Stutz, Swallows, Tobey, Tran, Tran,
  Vogt, Volin, Walker, Zolot, and Pino]{quanbtiuumH2}
S. A. Moses, C. H. Baldwin, M. S. Allman, R.~Ancona, L.~Ascarrunz,
  C.~Barnes, J.~Bartolotta, B.~Bjork, P.~Blanchard, M.~Bohn, J. G. Bohnet,
  N. C. Brown, N. Q. Burdick, W. C. Burton, S. L. Campbell, J. P.
  Campora, C.~Carron, J.~Chambers, J. W. Chan, Y. H. Chen, A.~Chernoguzov,
  E.~Chertkov, J.~Colina, J. P. Curtis, R.~Daniel, M.~DeCross, D.~Deen,
  C.~Delaney, J. M. Dreiling, C. T. Ertsgaard, J.~Esposito, B.~Estey,
  M.~Fabrikant, C.~Figgatt, C.~Foltz, M.~Foss-Feig, D.~Francois, J. P.
  Gaebler, T. M. Gatterman, C. N. Gilbreth, J.~Giles, E.~Glynn, A.~Hall,
  A. M. Hankin, A.~Hansen, D.~Hayes, B.~Higashi, I. M. Hoffman, B.~Horning,
  J. J. Hout, R.~Jacobs, J.~Johansen, L.~Jones, J.~Karcz, T.~Klein,
  P.~Lauria, P.~Lee, D.~Liefer, S. T. Lu, D.~Lucchetti, C.~Lytle, A.~Malm,
  M.~Matheny, B.~Mathewson, K.~Mayer, D. B. Miller, M.~Mills, B.~Neyenhuis,
  L.~Nugent, S.~Olson, J.~Parks, G. N. Price, Z.~Price, M.~Pugh, A.~Ransford,
  A. P. Reed, C.~Roman, M.~Rowe, C.~Ryan-Anderson, S.~Sanders, J.~Sedlacek,
  P.~Shevchuk, P.~Siegfried, T.~Skripka, B.~Spaun, R. T. Sprenkle, R. P.
  Stutz, M.~Swallows, R. I. Tobey, A.~Tran, T.~Tran, E.~Vogt, C.~Volin,
  J.~Walker, A. M. Zolot, and J. M. Pino.
\newblock A race-track trapped-ion quantum processor.
\newblock \emph{Physical Review X}, 13\penalty0 (4), December
  2023{\natexlab{c}}.
\newblock ISSN 2160-3308.
\newblock \doi{10.1103/physrevx.13.041052}.
\newblock URL \url{http://dx.doi.org/10.1103/PhysRevX.13.041052}.

\bibitem[Nie et~al.(2024{\natexlab{a}})Nie, Zi, and Sun]{Nie_2024}
Junhong Nie, Wei Zi, and Xiaoming Sun.
\newblock Quantum circuit for multi-qubit toffoli gate with optimal resource,
  2024{\natexlab{a}}.
\newblock URL \url{https://arxiv.org/abs/2402.05053}.

\bibitem[Nie et~al.(2024{\natexlab{b}})Nie, Zi, and Sun]{nie2024quantum}
Junhong Nie, Wei Zi, and Xiaoming Sun.
\newblock Quantum circuit for multi-qubit toffoli gate with optimal resource.
\newblock \emph{arXiv preprint arXiv:2402.05053}, 2024{\natexlab{b}}.

\bibitem[Norcia et~al.(2024)Norcia, Kim, Cairncross, Stone, Ryou, Jaffe, Brown,
  Barnes, Battaglino, Bohdanowicz, et~al.]{norcia2024iterative}
MA~Norcia, H~Kim, WB~Cairncross, M~Stone, A~Ryou, M~Jaffe, MO~Brown, K~Barnes,
  P~Battaglino, TC~Bohdanowicz, et~al.
\newblock Iterative assembly of 171 yb atom arrays with cavity-enhanced optical
  lattices.
\newblock \emph{PRX Quantum}, 5\penalty0 (3):\penalty0 030316, 2024.

\bibitem[Orts et~al.(2022)Orts, Ortega, and Garz{\'o}n]{orts2022studying}
Francisco Orts, Gloria Ortega, and Ester~M Garz{\'o}n.
\newblock Studying the cost of n-qubit toffoli gates.
\newblock In \emph{International Conference on Computational Science}, pages
  122--128. Springer, 2022.

\bibitem[Paetznick and Reichardt(2013)]{Paetznick_2013}
Adam Paetznick and Ben~W. Reichardt.
\newblock Universal fault-tolerant quantum computation with only transversal
  gates and error correction.
\newblock \emph{Physical Review Letters}, 111\penalty0 (9), August 2013.
\newblock ISSN 1079-7114.
\newblock \doi{10.1103/physrevlett.111.090505}.
\newblock URL \url{http://dx.doi.org/10.1103/PhysRevLett.111.090505}.

\bibitem[Satoh et~al.(2020)Satoh, Ohkura, and Van~Meter]{Satoh_2020}
Takahiko Satoh, Yasuhiro Ohkura, and Rodney Van~Meter.
\newblock Subdivided phase oracle for nisq search algorithms.
\newblock \emph{IEEE Transactions on Quantum Engineering}, 1:\penalty0 1–15,
  2020.
\newblock ISSN 2689-1808.
\newblock \doi{10.1109/tqe.2020.3012068}.
\newblock URL \url{http://dx.doi.org/10.1109/TQE.2020.3012068}.

\bibitem[Self et~al.(2024)Self, Benedetti, and Amaro]{iceberg}
Chris~N. Self, Marcello Benedetti, and David Amaro.
\newblock Protecting expressive circuits with a quantum error detection code.
\newblock \emph{Nature Physics}, 20\penalty0 (2):\penalty0 219–224, January
  2024.
\newblock ISSN 1745-2481.
\newblock \doi{10.1038/s41567-023-02282-2}.
\newblock URL \url{http://dx.doi.org/10.1038/s41567-023-02282-2}.

\bibitem[Sivarajah et~al.(2020)Sivarajah, Dilkes, Cowtan, Simmons, Edgington,
  and Duncan]{sivarajah2020t}
Seyon Sivarajah, Silas Dilkes, Alexander Cowtan, Will Simmons, Alec Edgington,
  and Ross Duncan.
\newblock t| ket>: a retargetable compiler for nisq devices.
\newblock \emph{Quantum Science and Technology}, 6\penalty0 (1):\penalty0
  014003, 2020.

\bibitem[Steane(1996)]{steane1996simple}
Andrew~M Steane.
\newblock Simple quantum error-correcting codes.
\newblock \emph{Physical Review A}, 54\penalty0 (6):\penalty0 4741, 1996.

\bibitem[Wang et~al.(2023)Wang, Simsek, Gatterman, Gerber, Gilmore, Gresh,
  Hewitt, Horst, Matheny, Mengle, Neyenhuis, and Criger]{wang2023faulttolerant}
Yang Wang, Selwyn Simsek, Thomas~M. Gatterman, Justin~A. Gerber, Kevin Gilmore,
  Dan Gresh, Nathan Hewitt, Chandler~V. Horst, Mitchell Matheny, Tanner Mengle,
  Brian Neyenhuis, and Ben Criger.
\newblock Fault-tolerant one-bit addition with the smallest interesting colour
  code, 2023.

\bibitem[Wang et~al.(2024)Wang, Simsek, Gatterman, Gerber, Gilmore, Gresh,
  Hewitt, Horst, Matheny, Mengle, et~al.]{wang2024fault}
Yang Wang, Selwyn Simsek, Thomas~M Gatterman, Justin~A Gerber, Kevin Gilmore,
  Dan Gresh, Nathan Hewitt, Chandler~V Horst, Mitchell Matheny, Tanner Mengle,
  et~al.
\newblock Fault-tolerant one-bit addition with the smallest interesting color
  code.
\newblock \emph{Science Advances}, 10\penalty0 (29):\penalty0 eado9024, 2024.

\bibitem[Yamamoto et~al.(2024)Yamamoto, Duffield, Kikuchi, and
  Mu{\~n}oz~Ramo]{yamamoto2024demonstrating}
Kentaro Yamamoto, Samuel Duffield, Yuta Kikuchi, and David Mu{\~n}oz~Ramo.
\newblock Demonstrating bayesian quantum phase estimation with quantum error
  detection.
\newblock \emph{Physical Review Research}, 6\penalty0 (1):\penalty0 013221,
  2024.

\end{thebibliography}

\appendix
\onecolumn

\section{Grover's Algorithm}\label{sec:grovers-algorithm}
\begin{figure}[tbp]
	\centering
	\includegraphics[width=0.7\linewidth]{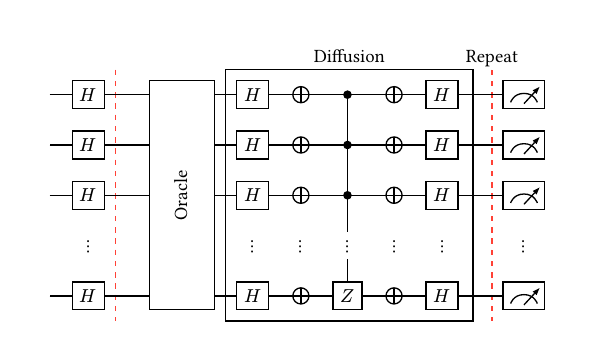}
	\caption{The general circuit for Grover's search.}
	\label{fig:grover-circ}
\end{figure}

Grover's algorithm or quantum search~\cite{grover1996fast} is a well-known quantum computing algorithm for increasing the probability of measuring from a set $\mathcal{M}$ of \textit{marked} qubit configurations $\ket{\mathbf{x}} \coloneqq \ket{x_0, \dots, x_n}$ with respect to a boolean oracle $f(\mathbf{x}) = \delta_{\mathbf{x} \in \mathcal{M}}$. Depicted in \cref{fig:grover-circ}, the core algorithm begins with the preparation of the $\ket{+}^{\otimes n}$ state on an $n$-qubit register via the $H^{\otimes n}$ operator, after which the following subroutine is repeated:
\begin{enumerate}
	\item \textbf{Marker}: $\ket{\mathbf{x}} \rightarrow (-1)^{f(\mathbf{x})} \ket{\mathbf{x}}$, mark target states with a phase of $-1$.
	\item \textbf{Diffusion}: $\ket{\mathbf{x}} \rightarrow (2 \ket{s}\bra{s} - I) \ket{\mathbf{x}}$, where $\ket{s} \coloneqq \frac{1}{\sqrt{2^n}} \sum_{\mathbf{x} = 0}^{2^n} \ket{\mathbf{x}}$. This step is generally implemented via $H^{\otimes n} X^{\otimes n} [C^{\otimes (n-1)} Z] X^{\otimes n} H^{\otimes n}$ as shown in \cref{fig:grover-circ}.
\end{enumerate}
Finally, the $n$-qubit register is measured in the computational $Z$-basis.
The probability of measuring a marked state after $k$ iterations with $N \coloneqq 2^n$ total states and $M \coloneqq |\mathcal{M}|$ marked states, where $M \leq N$, is given by:
\begin{equation}
	\mathbb{P}(\text{marked}) = \sin^2\left((2k+1) \arcsin\left(\sqrt{\frac{M}{N}}\right)\right)\label{eq:equation}
\end{equation}
It can then be shown that to maximize the probability of measuring a marked state, the optimal number of marker-diffusion iterations should be chosen as $\left\lfloor \frac{\pi}{4} \sqrt{\frac{N}{M}} - \frac{1}{2} \right\rfloor$~\cite{grover1996fast}, hence offering a quadratic speedup compared to a randomized classical sampling algorithm with success probability described by the negative hypergeometric distribution with mean $\mu = \frac{N+1}{M+1}$.

\section{Subroutines}\label{sec:subroutines}
\begin{algorithm}[H]
	\SetAlgoLined
	\DontPrintSemicolon
	\SetKwFunction{SyndromeMeasurement}{SyndromeMeasurement}

	\SetKwFunction{AddClassicalRegister}{AddClassicalRegister}
	\SetKwFunction{Barrier}{Barrier}
	\SetKwFunction{Hadamard}{H}
	\SetKwFunction{CNOT}{CX}
	\SetKwFunction{Measure}{Measure}
	\SetKwFunction{Reset}{Reset}

	\SetKwProg{Fn}{Function}{:}{}
	\Fn{\SyndromeMeasurement{$\Qubit$, $\Ancilla$}}{
		\tcp{$\Qubit$: main data register of size $n=k+2$}
		\tcp{$\Ancilla$: 2 ancillary qubits for error detection}

		$\mathcal{Z} \gets \text{\AddClassicalRegister}(1)$\;
		$\mathcal{X} \gets \text{\AddClassicalRegister}(1)$\;

		\text{\Hadamard}($\Ancilla_1$)\;

		\tcp{Initial CNOT operations}
		\ForEach{$(c, t, m) \in \{(1, 0, 0), (0, 0, 1), (1, 0, 1), (1, 1, 0)\}$}{
			\uIf{$m = 1$}{
				\text{\CNOT}($\Qubit_c$, $\Ancilla_t$)\;
			}
			\Else{
				\text{\CNOT}($\Ancilla_c$, $\Qubit_t$)\;
			}
		}

		\tcp{Main CNOT operations}
		\For{$i \gets 2$ \textbf{to} $k$ \textbf{step} $2$}{
			\ForEach{$(c, t, m) \in \{(1, i, 0), (i, 0, 1), (1, i+1, 0), (i+1, 0, 1)\}$}{
				\uIf{$m = 1$}{
					\text{\CNOT}($\Qubit_c$, $\Ancilla_t$)\;
				}
				\Else{
					\text{\CNOT}($\Ancilla_c$, $\Qubit_t$)\;
				}
			}
		}

		\tcp{Final CNOT operations}
		\ForEach{$(c, t, m) \in \{(1, -2, 0), (-2, 0, 1), (-1, 0, 1), (1, -1, 0)\}$}{
			\uIf{$m = 1$}{
				\text{\CNOT}($\Qubit_c$, $\Ancilla_t$)\;
			}
			\Else{
				\text{\CNOT}($\Ancilla_c$, $\Qubit_t$)\;
			}
		}

		\text{\Hadamard}($\Ancilla_1$)\;
		\text{\Measure}($\Ancilla_0\to \mathcal{Z}$)\;
		\text{\Measure}($\Ancilla_1\to \mathcal{X}$)\;
		\text{\Reset}($\Ancilla$)\;

		\KwRet{$\mathcal{Z}$, $\mathcal{X}$}\;
	}
	\caption{Syndrome Measurement Subroutine with Qubit ($\Qubit$) and Ancilla ($\Ancilla$) Registers}\label{alg:algorithm2}
\end{algorithm}

\begin{algorithm}[H]
	\SetAlgoLined
	\DontPrintSemicolon
	\SetKwFunction{PrepareZeroState}{PrepareZeroState}

	\SetKwFunction{Barrier}{Barrier}
	\SetKwFunction{Hadamard}{H}
	\SetKwFunction{CNOT}{CX}
	\SetKwFunction{Measure}{Measure}
	\SetKwFunction{Reset}{Reset}

	\SetKwProg{Fn}{Function}{:}{}
	\Fn{\PrepareZeroState{$\Qubit$, $\Ancilla$}}{
		\tcp{$\Qubit$: main data register of size $n=k+2$}
		\tcp{$\Ancilla$: 1 ancillary qubit for flagging and measurement}

		$\mathcal{F} \gets \text{\AddClassicalRegister}(1)$\;

		\tcp{Construct GHZ state on $\Qubit$}
		\text{\Hadamard}($\Qubit_0$)\;
		\For{$i \gets 1$ \textbf{to} $n-1$}{
			\text{\CNOT}($\Qubit_{i-1}$, $\Qubit_i$)\;
		}

		\tcp{Flag potential high weight errors onto $\Ancilla_0$}
		\text{\CNOT}($\Qubit_0$, $\Ancilla_0$)\;
		\text{\CNOT}($\Qubit_{n-1}$, $\Ancilla_0$)\;

		\tcp{Measure the flag qubit}
		\text{\Measure}($\Ancilla_0 \to \mathcal{F}$)\;

		\text{\Reset}($\Ancilla_0$)\;

		\KwRet{$\mathcal{F}$}\;
	}
	\caption{Zero State Preparation Subroutine with Qubit ($\Qubit$) and Ancilla ($\Ancilla$) Registers}
	\label{alg:ft-zero-state}
\end{algorithm}

\begin{algorithm}[H]
	\SetAlgoLined
	\DontPrintSemicolon
	\SetKwFunction{MeasureAll}{MeasureAll}

	\SetKwFunction{Hadamard}{H}
	\SetKwFunction{CNOT}{CX}
	\SetKwFunction{Measure}{Measure}
	\SetKwFunction{Barrier}{Barrier}

	\SetKwProg{Fn}{Function}{:}{}
	\Fn{\MeasureAll{$\Qubit$, $\Ancilla$}}{
		\tcp{$\Qubit$: main data register of size $n$}
		\tcp{$\Ancilla$: 2 ancillary qubits for parity check and syndrome detection}

		$\mathcal{C} \gets \text{\AddClassicalRegister}(n)$\;
		$\mathcal{F} \gets \text{\AddClassicalRegister}(1)$\;
		$\mathcal{X} \gets \text{\AddClassicalRegister}(1)$\;

		\text{\Hadamard}($A_0$)\;
		\text{\CNOT}($A_0$, $Q_{n-1}$)\;
		\text{\CNOT}($A_0$, $A_1$)\;

		\For{$i \gets 1$ \textbf{to} $n-2$}{
			\text{\CNOT}($a_0$, $\Qubit_i$)\;
		}

		\text{\CNOT}($a_0$, $a_1$)\;
		\text{\CNOT}($a_0$, $b$)\;

		\text{\Hadamard}($a_0$)\;

		\tcp{Z basis measurement for data register}
		\text{\Measure}($\Qubit \to \mathcal{C}_\text{data}$)\;

		\tcp{Measure $X$ syndrome and final flag off ancillas}
		\text{\Measure}($a_0 \to \mathcal{X}$)\;
		\text{\Measure}($a_1 \to \mathcal{F}$)\;

		\KwRet{$\mathcal{C}$, $\mathcal{X}$, $\mathcal{F}$}\;
	}
	\caption{Destructive X Syndrome and Z Measurement Subroutine with Qubit ($\Qubit$) and Ancilla ($\Ancilla$) Registers}\label{alg:algorithm}
\end{algorithm}

\begin{algorithm}[H]

	\SetAlgoLined
	\DontPrintSemicolon
	\SetKwFunction{LogicalPauli}{LogicalPauli}

	\SetKwFunction{CNOT}{CX}
	\SetKwFunction{Hadamard}{H}
	\SetKwFunction{XGate}{X}
	\SetKwFunction{ZGate}{Z}
	\SetKwFunction{IncrementOpCount}{IncrementOpCount}

	\SetKwProg{Fn}{Function}{:}{}
	\Fn{\LogicalPauli{$\Qubit$, $\beta$, $\tau$}}{
	\tcp{$\Qubit$: main data register of size $n=k+2$}
	\tcp{$\beta$: a list of qubits to apply the Pauli operator to}
	\tcp{$\tau$: type of Pauli operators $X$ or $Z$}
	\smallbreak\smallbreak

	\tcp{Shift indices up by 1}
	$\beta \gets [x + 1 \text{ for } x \in \beta]$\;

	\If{$\texttt{length}(bits)\ \% \ 2 \neq 0$}{
	\If{$\tau = z$}{
	$\beta \gets \beta + [n - 1]$\;
	}
	\Else{
		$\beta \gets [0] + \beta$\;
	}
	}

	\tcp{Take the complement to find a smaller representation}
	\If{$\texttt{length}(\beta) > n / 2$}{
		$\beta \gets [i \text{ for } i \in \texttt{range}(n) \text{ if } i \notin \beta]$\;
	}

	\tcp{Apply the corresponding physical Pauli operator}
	\uIf{$\tau = z$}{
		\text{\ZGate}($\Qubit_{\beta}$)\;
	}
	\Else{
		\text{\XGate}($\Qubit_{\beta}$)\;
	}
	}
	\caption{Logical Pauli Operator (X or Z) Subroutine}
	\label{alg:logical-paulis}
\end{algorithm}

\section{Multi-controlled Toffoli Gate Implementations}\label{sec:multi-controlled-toffoli-gate-implementations}
We explain the computation of the 1-qubit $U$ gate count, $\boldsymbol{\#}_U = 16k - 18$, the 2-qubit $CX$ gate count, $\boldsymbol{\#}_{CX} = 16k - 24$, and the circuit depth, $\boldsymbol{D} = 32k - 83$.

Constructions from~\cite{Claudon_2024} fit $\boldsymbol{D} \approx 64 \log(k)^3 - 263$ and $\boldsymbol{\#}_{CX} \approx 499k - 80557$, evaluated at the points where $k \in \{2^j \,|\, j \in \{6, \dots, 12\}\}$.

\begin{figure*}[h!]
	\centering
	\includegraphics[width=0.9\textwidth]{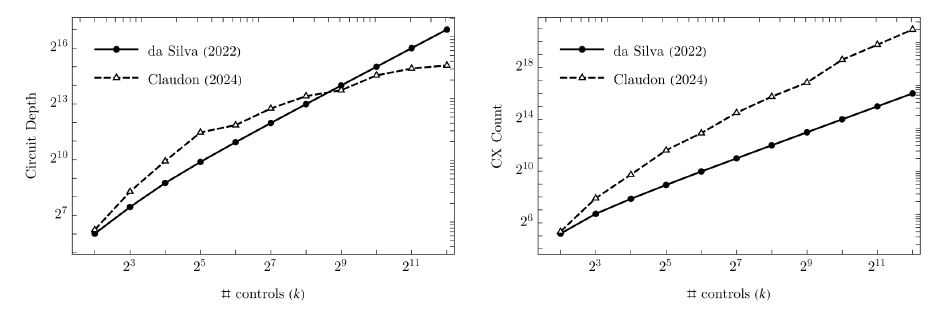}
	\caption{Comparison in 2-qubit gate counts between \cite{Claudon_2024} and \cite{PhysRevA.106.042602} for $C^{\otimes k}X$ decomposition.}
\end{figure*}

The decompositions from~\cite{Claudon_2024} also require considerably more time to compute compared to~\cite{PhysRevA.106.042602}, with times of 255.23s versus 13.95s, respectively, on an Apple M1 chip with 16GB of memory.

\section{Syndrome Experiments}\label{sec:syndrome-experiments}
\subsection{Toy Model with Noisy Identity Gates}\label{subsec:toy-model-with-noisy-identity-gates}

\begin{figure*}[ht]
	\centering
	\includegraphics[width=\textwidth]{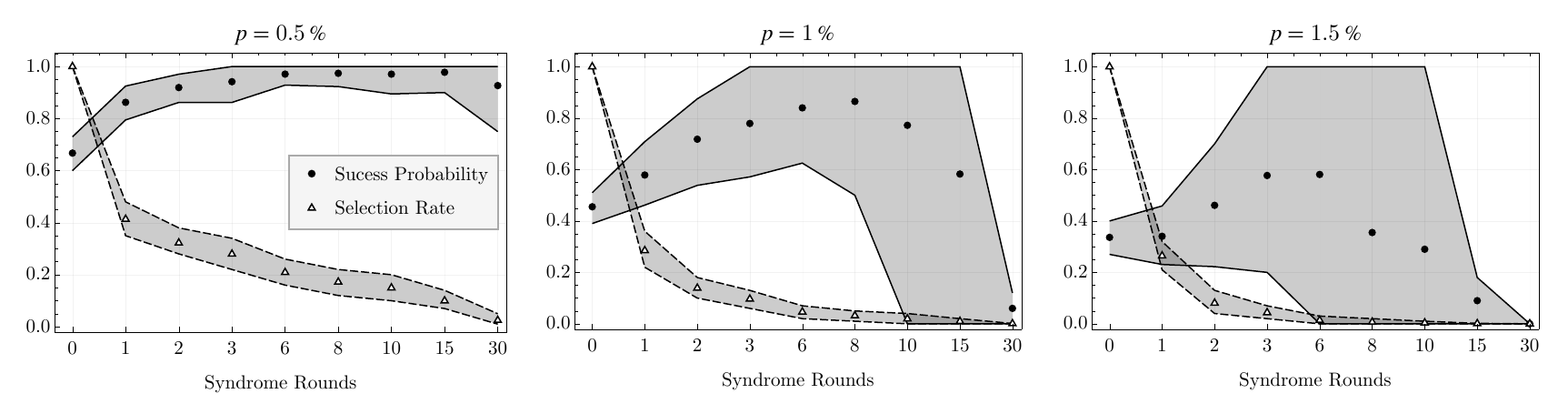}
	\caption{\textit{Syndrome scheduling vs algorithmic performance after post-selection.} Comparing the success 
    probability and post-selection rate of a noisy identity circuit over various error-detecting strategies under the 
    circuit-level noise model. Using both a non-fault-tolerant benchmark (syndrome rounds $=0$) and the proposed 
    fault-tolerant approach on the $qcode(n, n-2, 2)$, the frequency of syndrome extraction rounds is increased at 
    multiple noise levels $p=0.5\%$ (left), $1\%$ (center), and $1.5\%$ (right). Full experimental details are shown 
    in \ref{sec:syndrome-experiments}.}
	\label{fig:syndromes}
\end{figure*}

\begin{figure*}[h]
	\centering
	\begin{quantikz}[background color=black!5!white]
		\lstick[6, nwires={0}]{$\ket{0}^{\otimes 6}$} & \gate[6, nwires={0}]{\text{FT-Prepare}} & \gate{\tilde{I}} & \ \ldots \  & \gate[6, nwires={0}, disable auto
			height]{\verticaltext{Syndrome}} & \gate{\tilde{I}} & \ \ldots \  & \gate[6, nwires={0}]{\text{FT-Measure}} \\
		\lstick{} & & \gate{\tilde{I}} & \ \ldots \  & & \gate{\tilde{I}} & \ \ldots \  &  \\
		\lstick{} & & \gate{\tilde{I}} & \ \ldots \  & & \gate{\tilde{I}} & \ \ldots \  &  \\
		\lstick{} & & \gate{\tilde{I}} & \ \ldots \  & & \gate{\tilde{I}} & \ \ldots \  &  \\
		\lstick{} & & \gate{\tilde{I}} & \ \ldots \  & & \gate{\tilde{I}} & \ \ldots \  &  \\
		\lstick{} & & \gate{\tilde{I}} & \ \ldots \  & & \gate{\tilde{I}} & \ \ldots \  &
	\end{quantikz}
	\caption{Overview of syndrome scheduling experimental setup. After fault-tolerant state preparation, a sequence of 
    noisy identity gates $\tilde{I}$ and syndrome measurements proceeds. The syndrome scheduling rate is the main 
    experimental parameter, determining how many rounds of noisy identity gates pass before a syndrome round is 
    initiated. The final syndrome measurement destructively measures all qubits to compute $Z^{\otimes n}$ and measures
    $X^{\otimes n}$ onto an ancilla qubit using one flag.}
	\label{fig:syndrome-circuit}
\end{figure*}
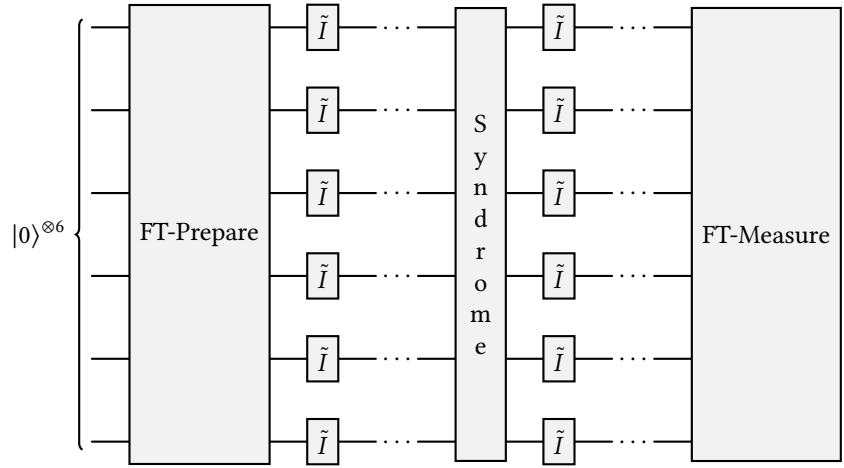

Summary of Experimental Details for Syndrome Scheduling Experiments (Referenced in \autoref{fig:syndrome-circuit}):

\begin{itemize}
	\item \textbf{Qubit Encoding:} Experiments were conducted using 4 logical qubits encoded with the $\qcode{6}{4}{2}$ code.

	\item \textbf{Noise Model:} A circuit-level noise model as described in \ref{sec:evaluation} was applied with
    error rates of $p = 0.5\%$, $1\%$, and $1.5\%$.

	\item \textbf{Circuit Operations:} Each qubit underwent 30 rounds of noisy identity gates between state preparation 
    and final measurements.

	\item \textbf{Syndrome Rounds:} The number of syndrome rounds varied across the following values: $r = \{0, 1, 2, 3, 6, 8, 10, 15, 30\}$.
	      \begin{itemize}
		      \item $r = 0$ corresponds to a non-fault-tolerant circuit using 4 physical qubits and no additional state 
              preparation overhead.
		      \item $r = 1$ corresponds to a single syndrome round applied at the end of the circuit, utilizing an 
              additional flag qubit as described in \cite{iceberg}.
		      \item $r > 1$ indicates that syndrome circuits were inserted after every ${1, 2, 3, 4, 5, 10, 15}$ rounds 
              of noise.
	      \end{itemize}

	\item \textbf{Post Selection:} Runs were post-selected out based on negative syndrome readings or triggered flag 
    qubits at any point in the circuit.

	\item \textbf{Fault Tolerance Encoding:} Fault-tolerant circuits began with the fault-tolerant GHZ state preparation
    circuit, as illustrated in \cite{iceberg}.

	\item \textbf{Shot Count:} Each circuit was executed for 100 shots.

	\item \textbf{Confidence Intervals:} Confidence intervals were calculated as the 80\% centered range around the mean
    from the empirical cumulative distribution function (CDF) over 100 runs.
\end{itemize}

\clearpage
\section{Experiment Results}\label{sec:experiment-results}
We present simulation results for all configurations of Grover's algorithm using \(\qcode{n}{n-2}{2}\) codes, covering
qubit counts \( n \in \{6, 8, 10, 12\} \), Grover iterations \( k \in \{1, 2, \dots, 4\} \), and noise rates \( p \)
ranging from \( 0.0125 \) to \( 0.8 \) in increments of \( 0.0125 \).
\begin{figure}[H]
    \centering
	\begin{subfigure}{0.87\textwidth}
		\includegraphics[width=1.0\linewidth]{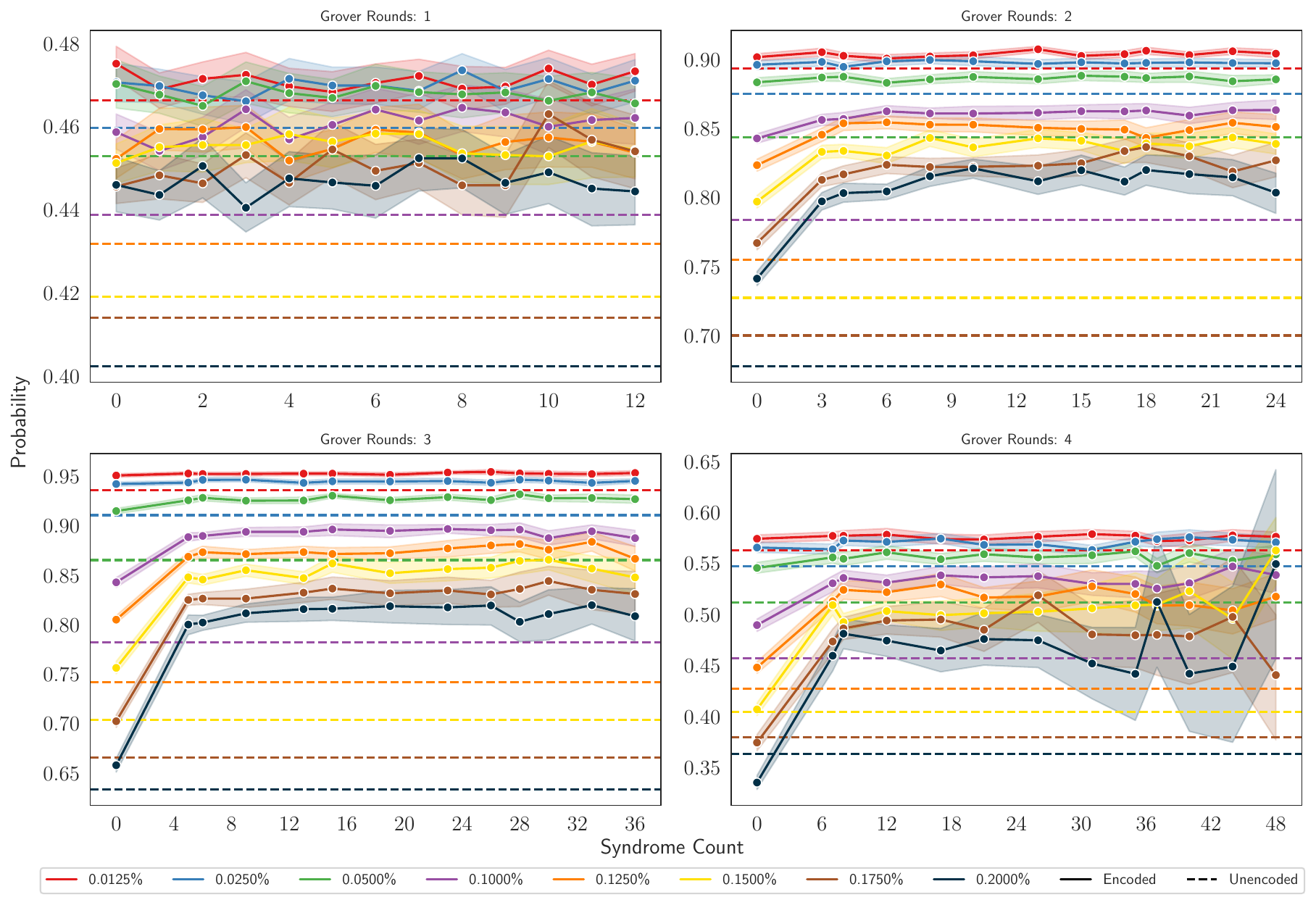}
	\end{subfigure}
    \vfill
	\begin{subfigure}{0.87\textwidth}
		\includegraphics[width=1.0\linewidth]{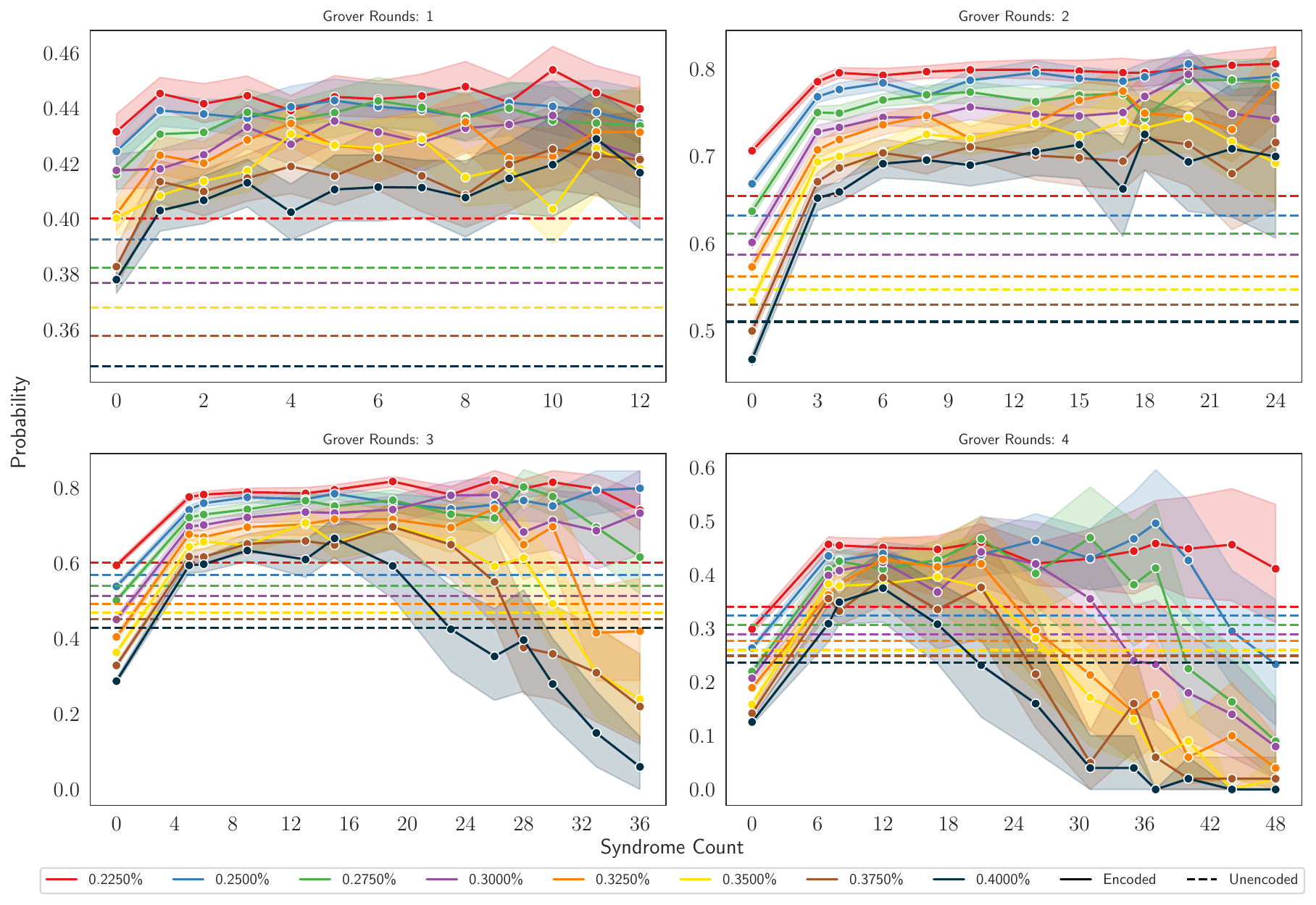}
	\end{subfigure}
    \caption{$\qcode{6}{4}{2}$}
\end{figure}
\begin{figure}[H]
    \centering
	\begin{subfigure}{1.0\textwidth}
		\includegraphics[width=1.0\linewidth]{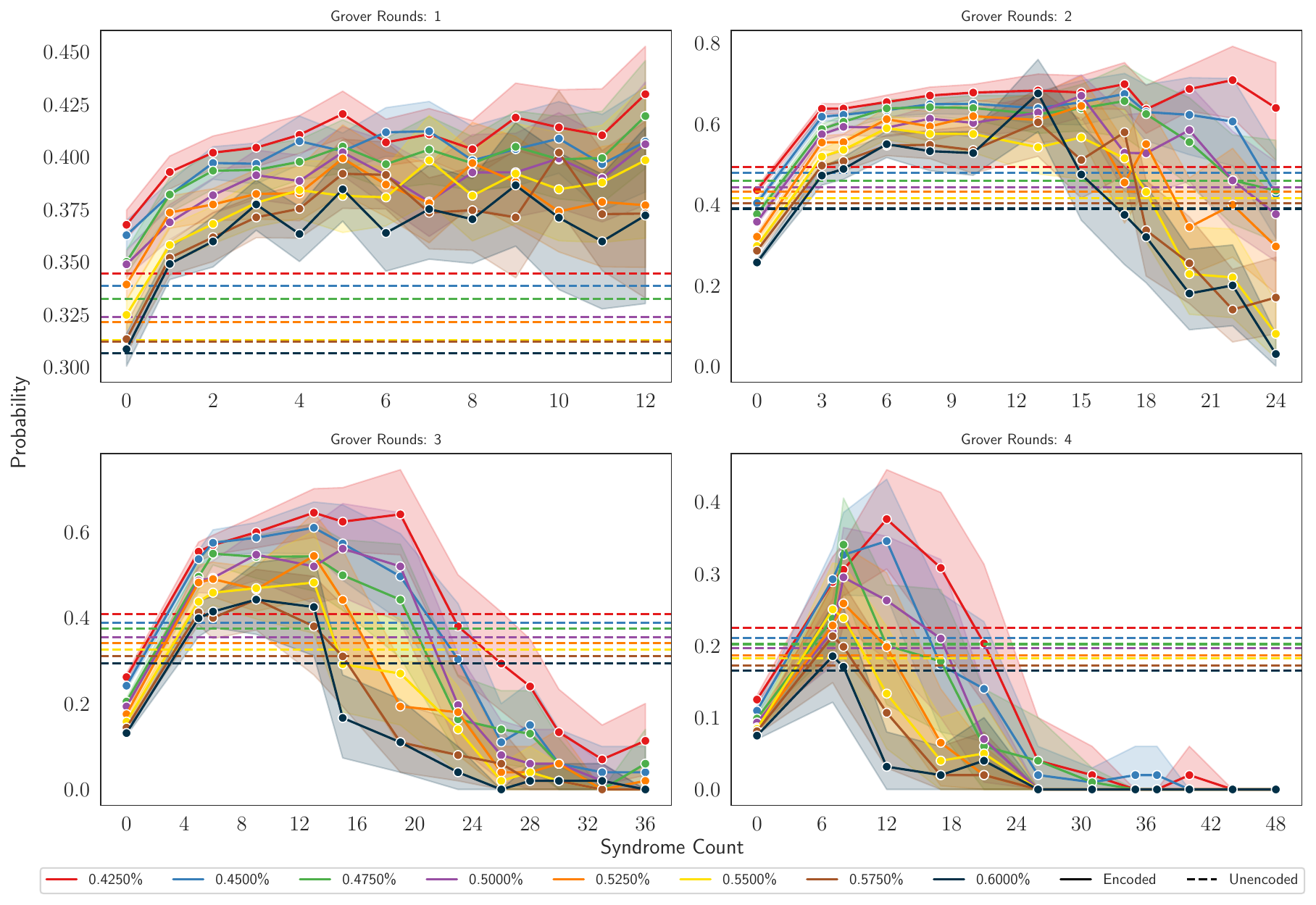}
	\end{subfigure}
    \vfill
	\begin{subfigure}{1.0\textwidth}
		\includegraphics[width=1.0\linewidth]{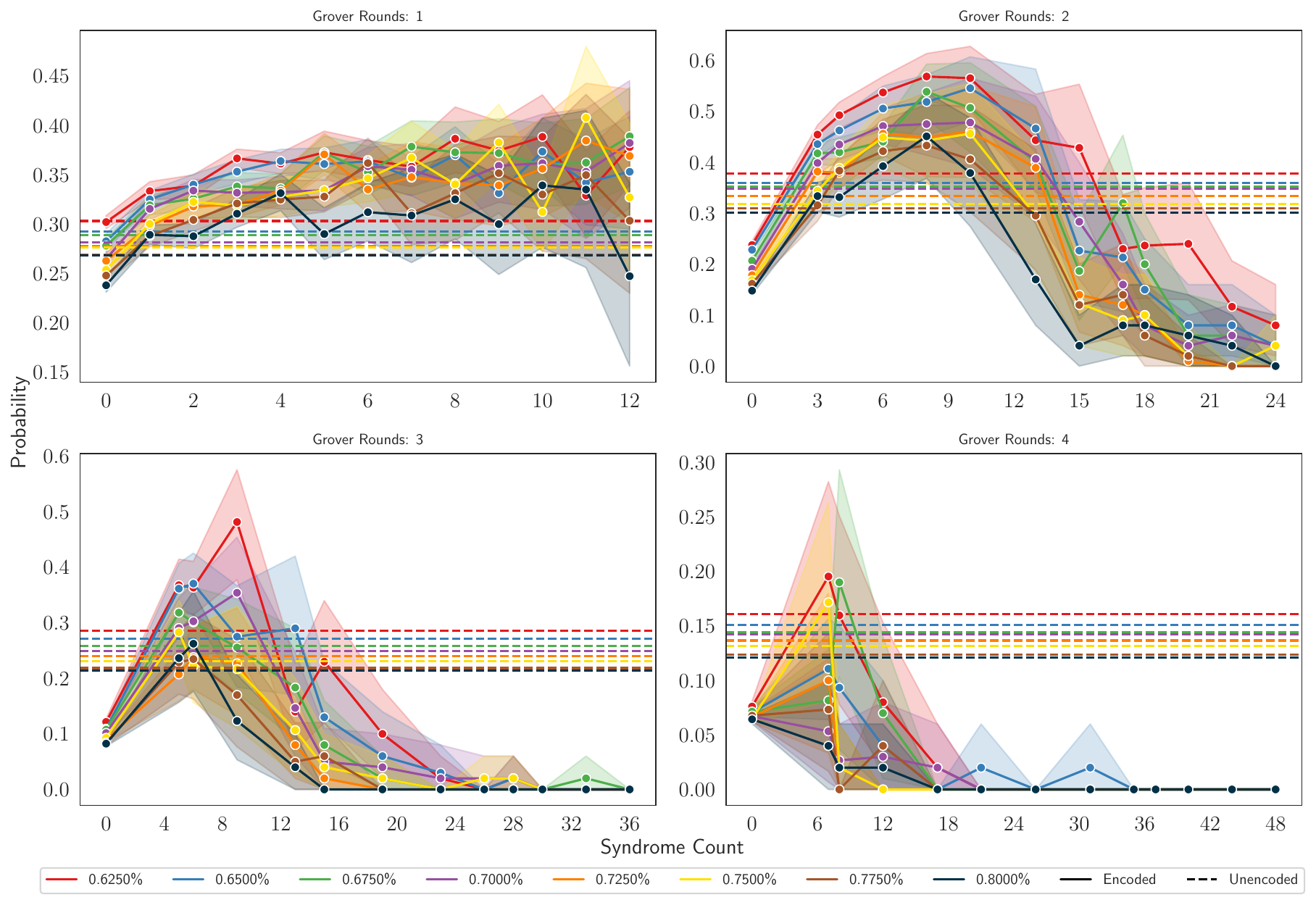}
	\end{subfigure}
    \caption{$\qcode{6}{4}{2}$}
\end{figure}
\begin{figure}[H]
    \centering
	\begin{subfigure}{1.0\textwidth}
		\includegraphics[width=1.0\linewidth]{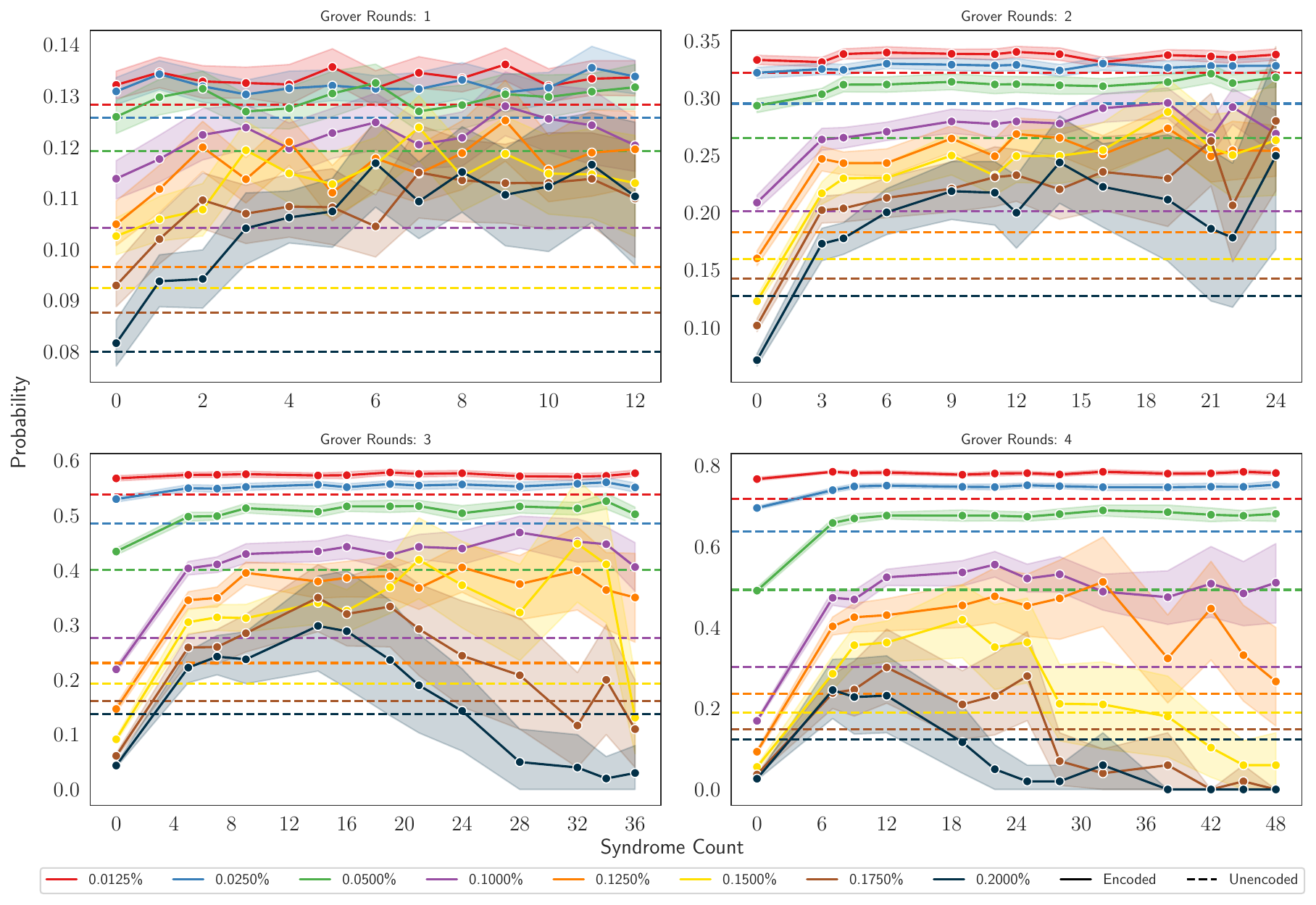}
	\end{subfigure}
    \vfill
	\begin{subfigure}{1.0\textwidth}
		\includegraphics[width=1.0\linewidth]{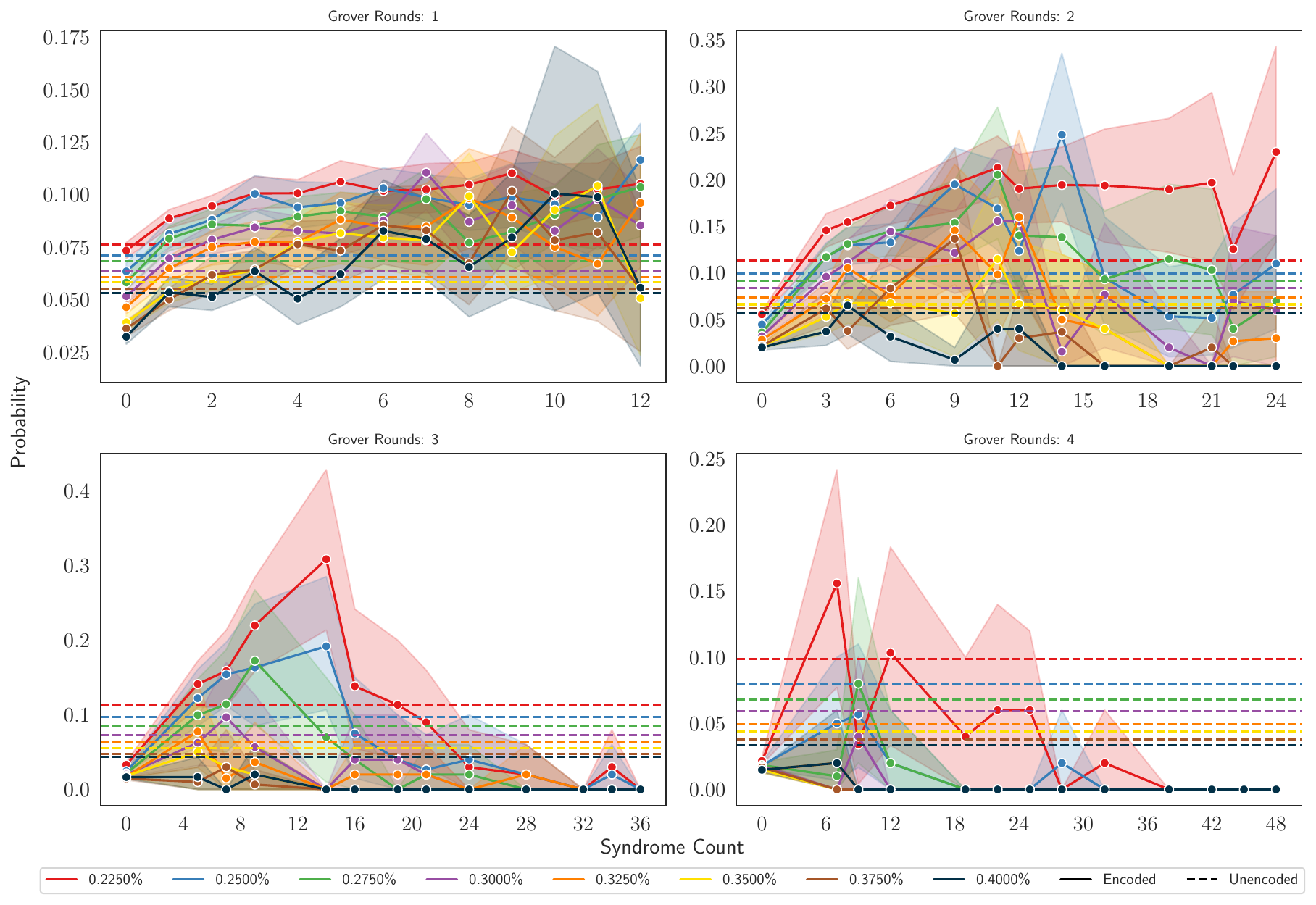}
	\end{subfigure}
    \caption{$\qcode{8}{6}{2}$}
\end{figure}
\begin{figure}[H]
    \centering
	\begin{subfigure}{1.0\textwidth}
		\includegraphics[width=1.0\linewidth]{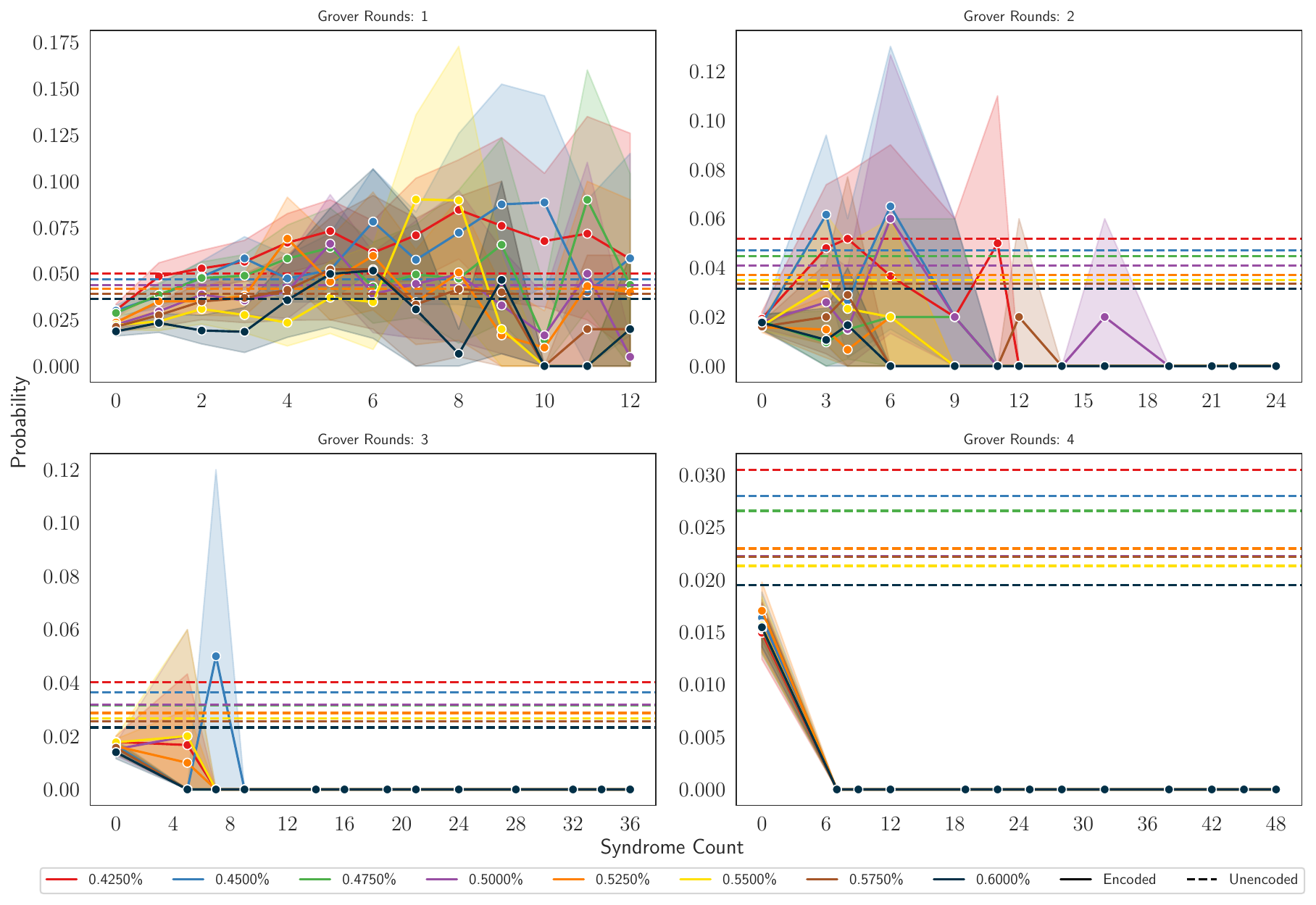}
	\end{subfigure}
    \vfill
	\begin{subfigure}{1.0\textwidth}
		\includegraphics[width=1.0\linewidth]{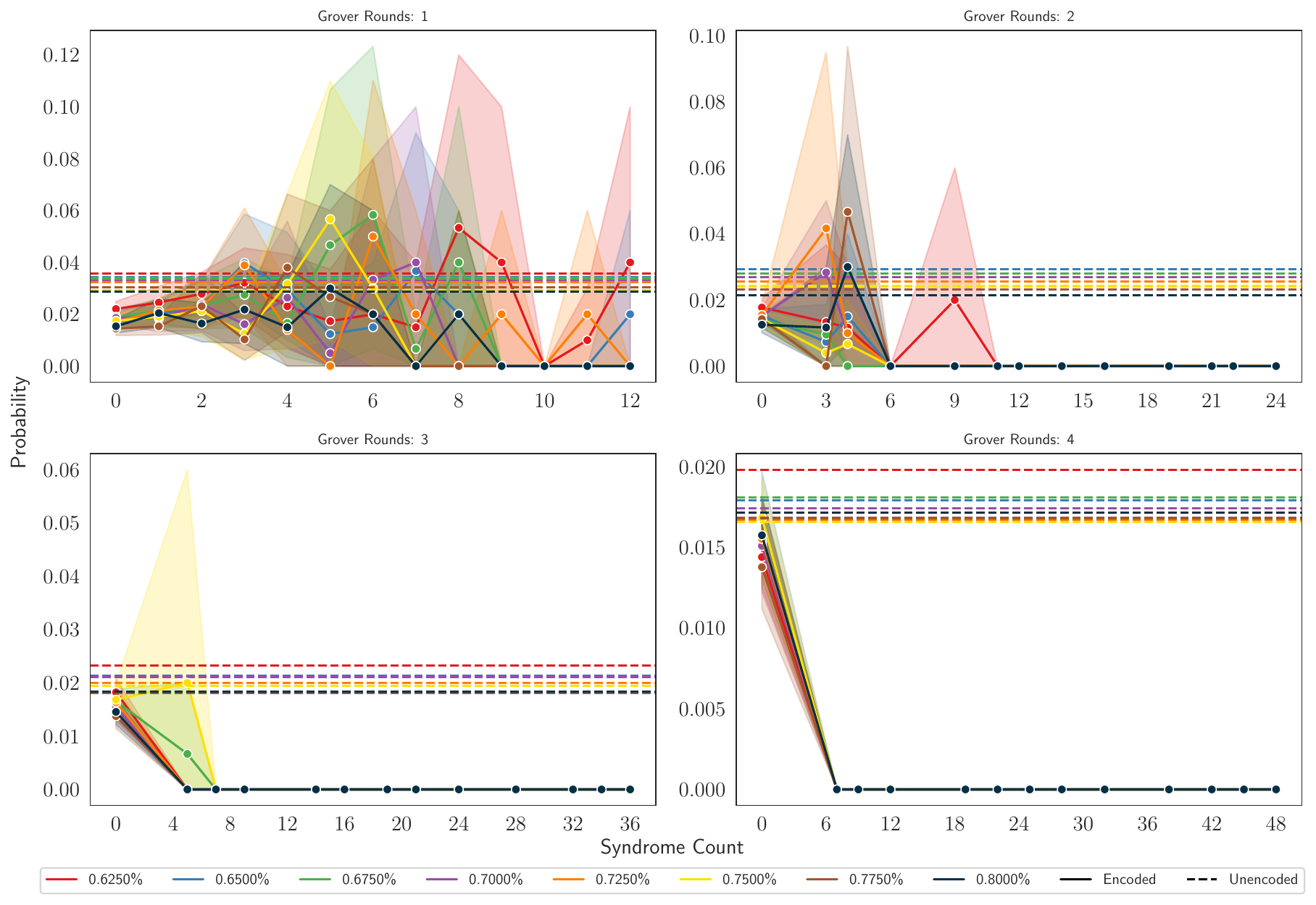}
	\end{subfigure}
    \caption{$\qcode{8}{6}{2}$}
\end{figure}

\begin{figure}[H]
	\begin{subfigure}{1.0\textwidth}
		\includegraphics[width=1.0\linewidth]{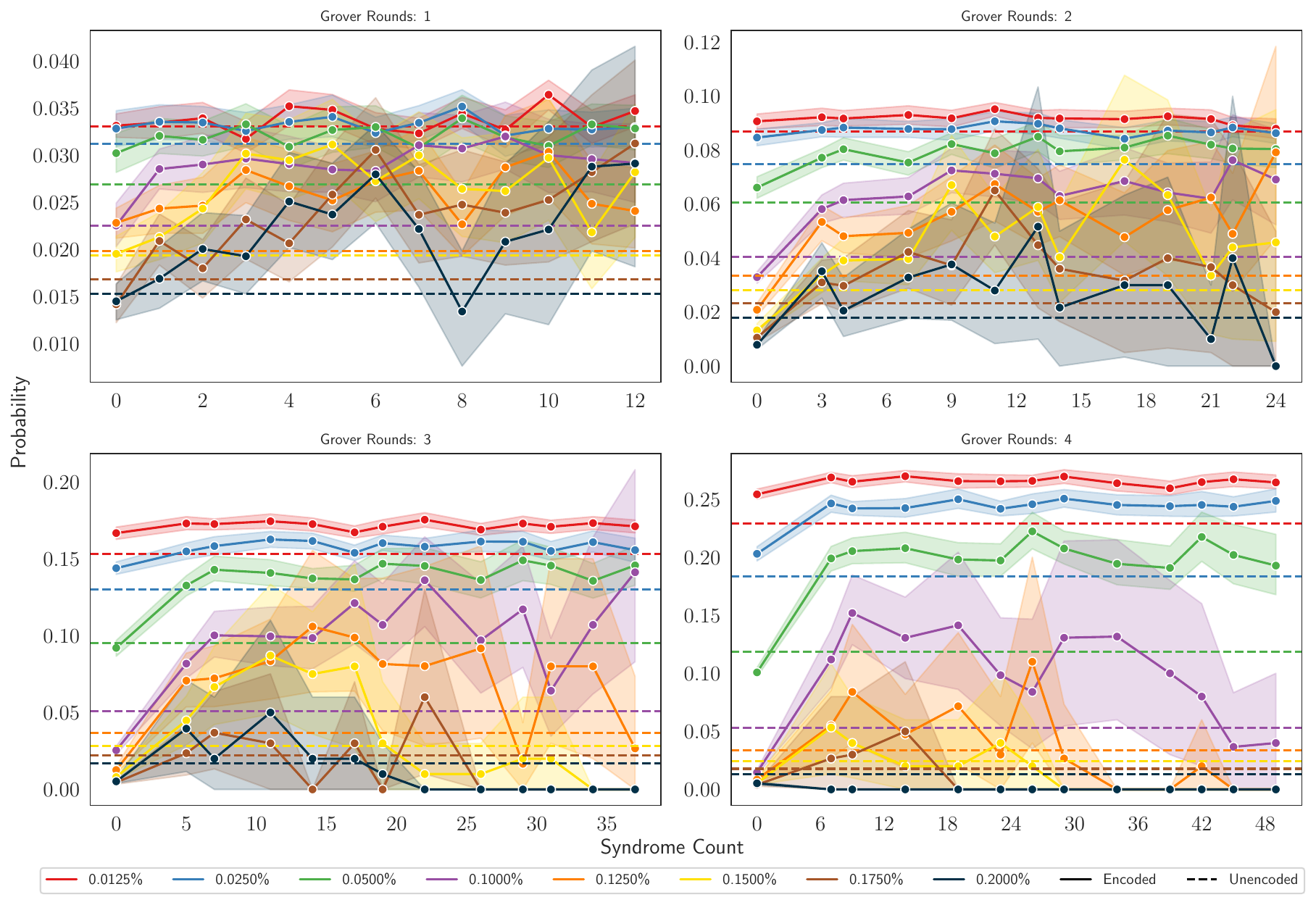}
	\end{subfigure}
    \vfill
	\begin{subfigure}{1.0\textwidth}
		\includegraphics[width=1.0\linewidth]{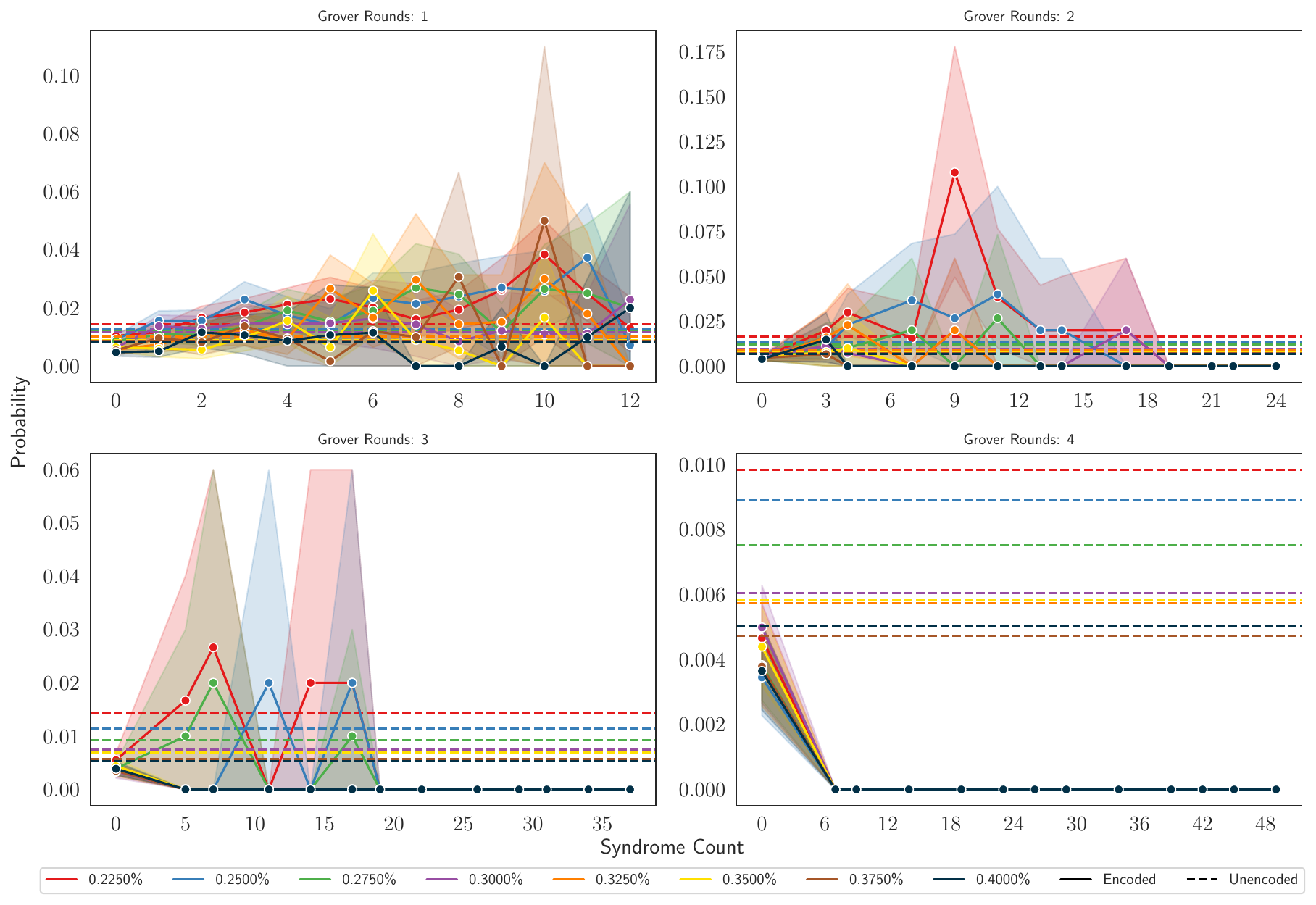}
	\end{subfigure}
    \caption{$\qcode{10}{8}{2}$}
\end{figure}
\begin{figure}[H]
	\begin{subfigure}{1.0\textwidth}
		\includegraphics[width=1.0\linewidth]{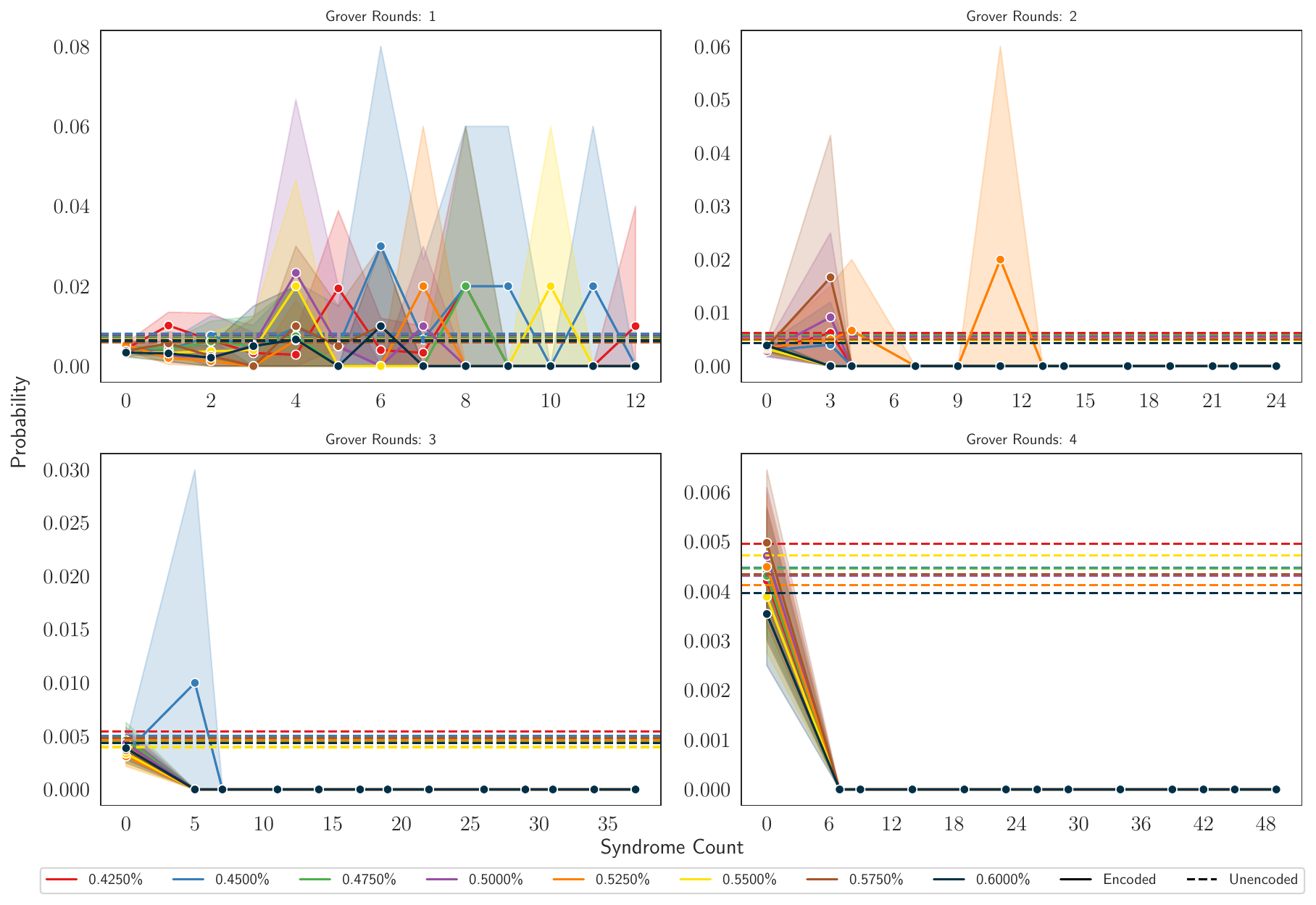}
	\end{subfigure}
    \vfill
	\begin{subfigure}{1.0\textwidth}
		\includegraphics[width=1.0\linewidth]{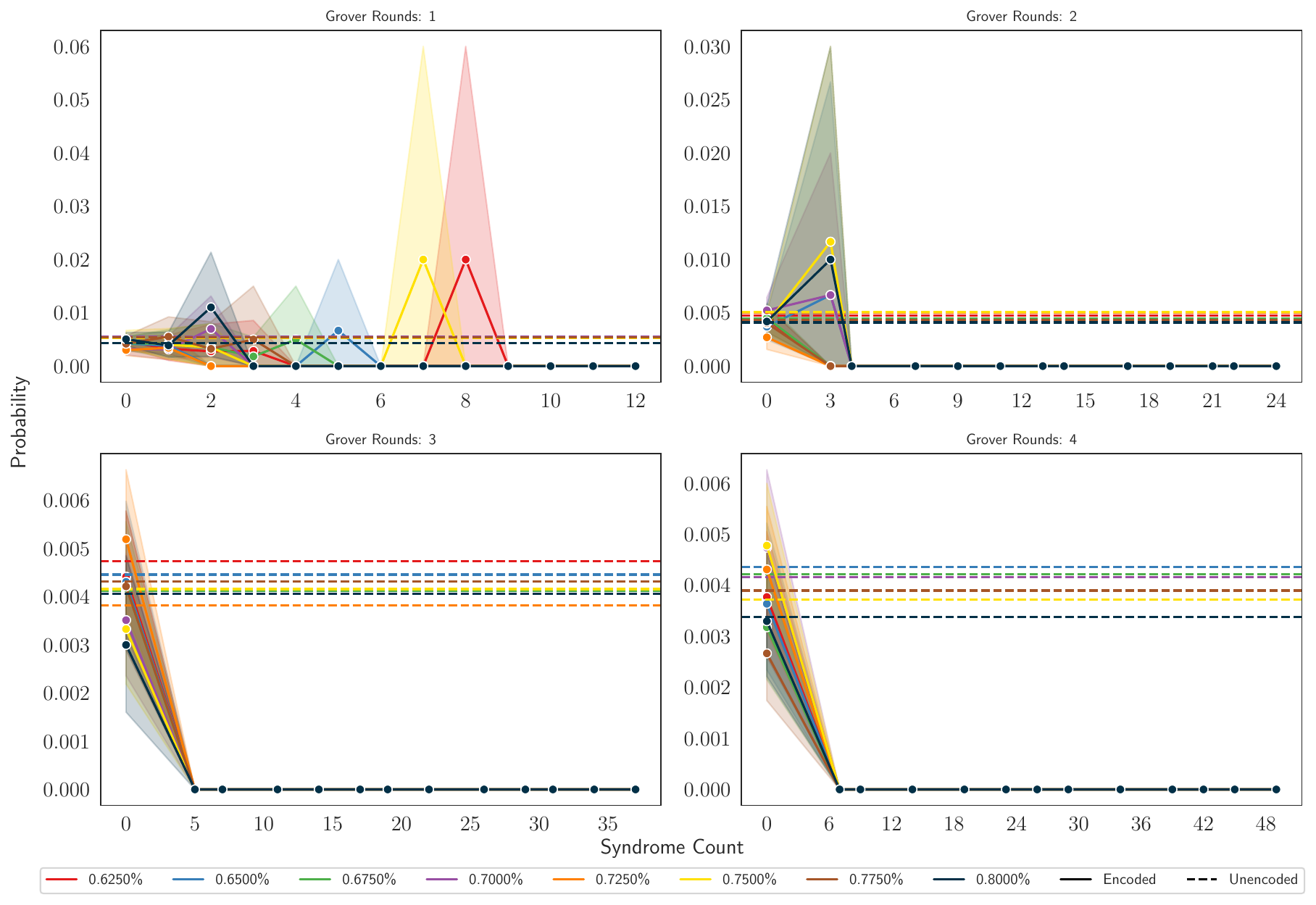}
	\end{subfigure}
    \caption{$\qcode{10}{8}{2}$}
\end{figure}

\begin{figure}[H]
	\begin{subfigure}{1.0\textwidth}
		\includegraphics[width=1.0\linewidth]{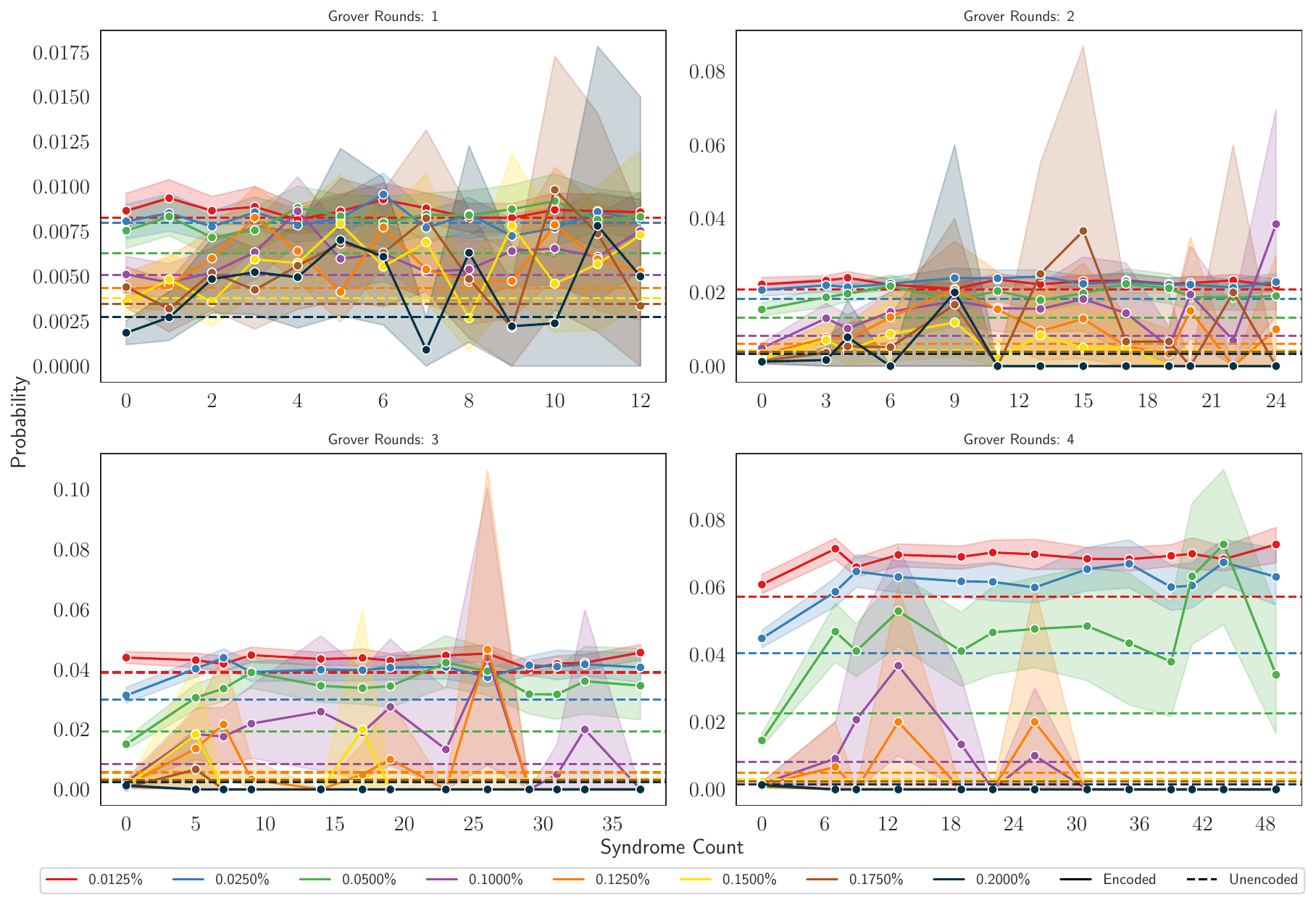}
	\end{subfigure}
    \vfill
	\begin{subfigure}{1.0\textwidth}
		\includegraphics[width=1.0\linewidth]{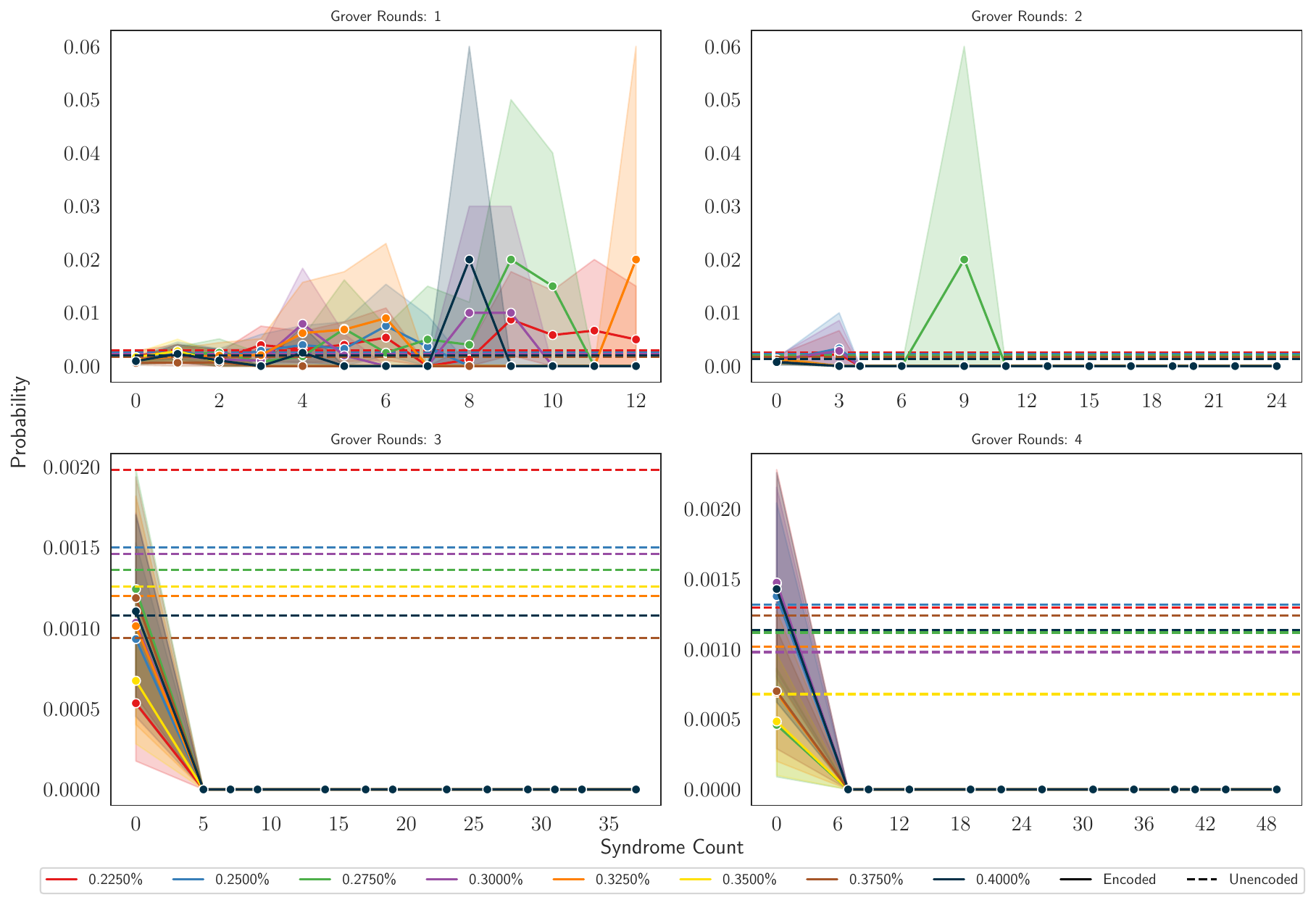}
	\end{subfigure}
    \caption{$\qcode{12}{10}{2}$}
    \label{app:12_10_2}
\end{figure}
\begin{figure}[H]
	\begin{subfigure}{1.0\textwidth}
		\includegraphics[width=1.0\linewidth]{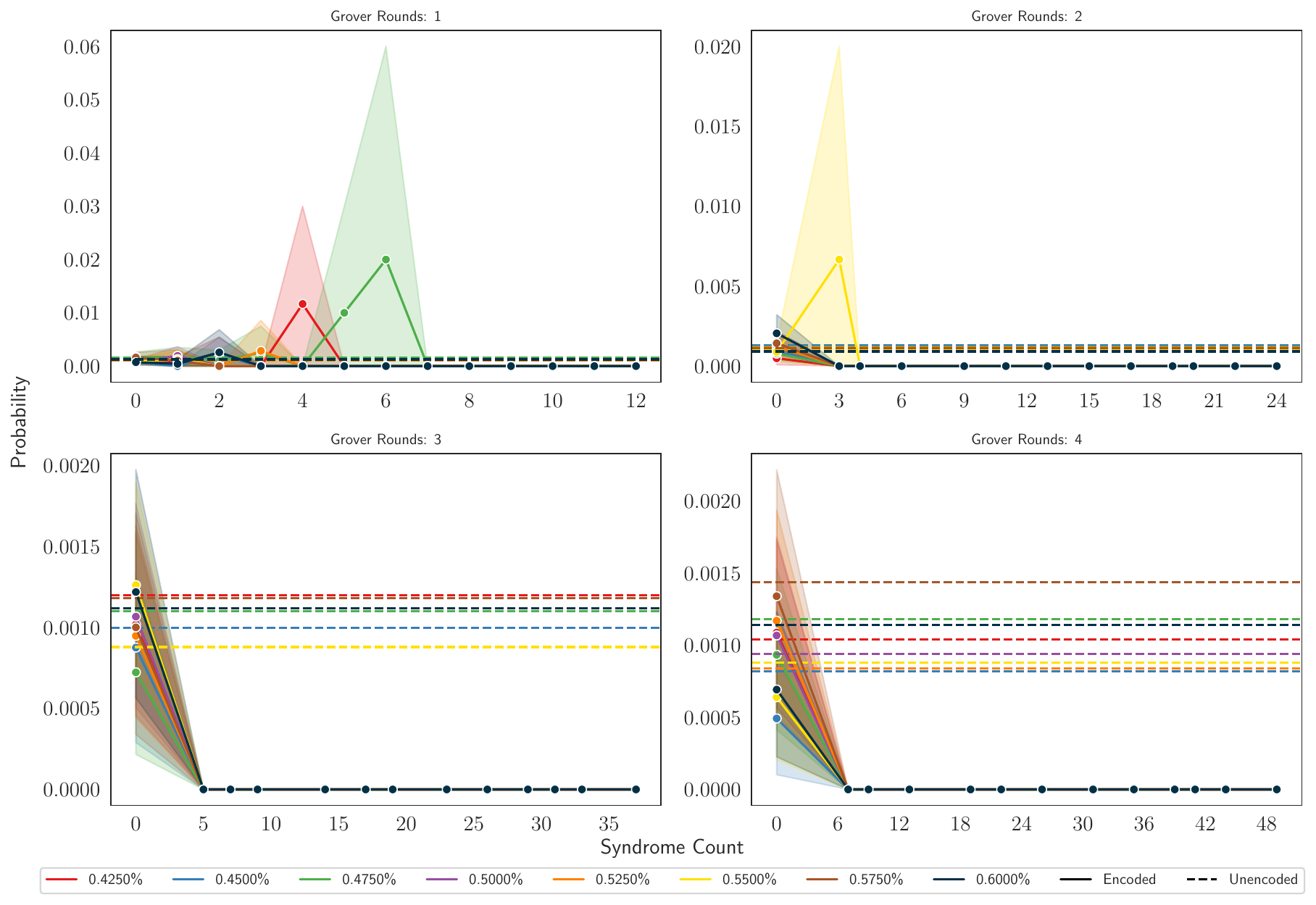}
	\end{subfigure}
    \vfill
	\begin{subfigure}{1.0\textwidth}
		\includegraphics[width=1.0\linewidth]{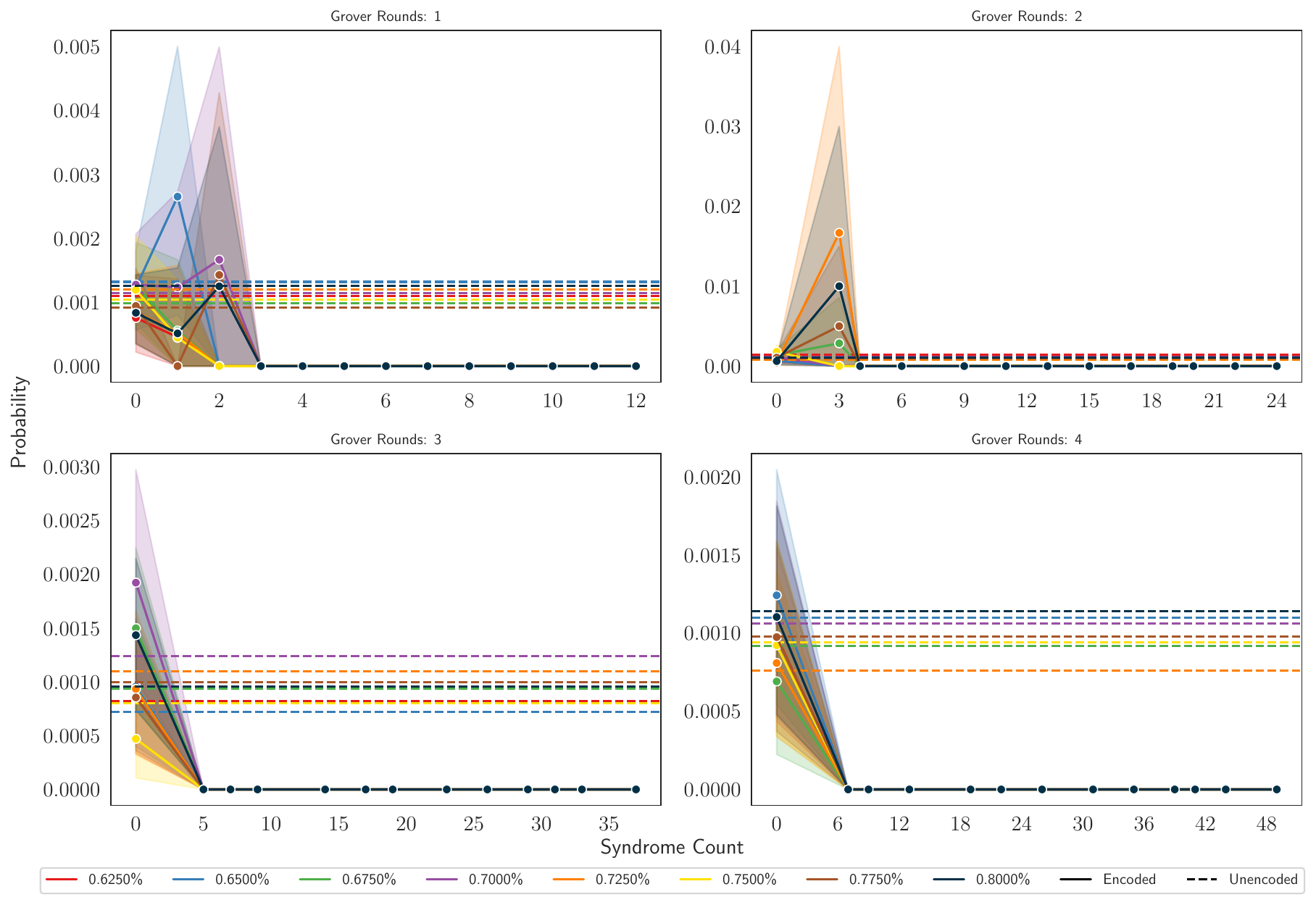}
	\end{subfigure}
    \caption{$\qcode{12}{10}{2}$}
\end{figure}

\end{document}